\documentclass[10pt, journal]{IEEEtran}
\IEEEoverridecommandlockouts
\usepackage[utf8]{inputenc}
\usepackage{amsthm}
\usepackage{amssymb}

\usepackage{multirow}
\usepackage{balance}
\usepackage{color,soul}
\usepackage{pgfplots}
\pgfplotsset{compat=1.18} % Set the compatibility level for pgfplots
\usepackage{physics}

\usepackage{algorithm}
 \usepackage{algpseudocode}
 \usepackage{algorithmicx}
\newtheorem{theorem}{Theorem}
\newtheorem{definition}{Definition}
\newtheorem{lemma}{Lemma}
\usepackage{siunitx}
\usepackage{flushend}

\usepackage[justification=centering]{caption}
\usepackage{titlesec}
\usepackage{hyperref}

\renewcommand{\qedsymbol}{\blacksquare}

\graphicspath{{/supplements/}}
\IEEEaftertitletext{\vspace{-25pt}}
\titlespacing*{\section}{0pt}{4pt}{2pt}
\titlespacing*{\subsection}{0pt}{2pt}{1pt}
\titlespacing*{\subsubsection}{0pt}{1pt}{1pt}

\begin{document}

\title{\fontsize{22pt}{24pt}\selectfont Resource Heterogeneity-Aware and Utilization-Enhanced Scheduling for Deep Learning Clusters \textsuperscript{\normalsize \(\dagger\)}\footnotemark
}
\author{%
    Abeda Sultana\textsuperscript{1}, Nabin Pakka\textsuperscript{1}, Fei Xu\textsuperscript{2}, Xu Yuan\textsuperscript{3}, Li Chen\textsuperscript{1}, and Nian-Feng Tzeng\textsuperscript{1}\\
    {\small % Specify font size for universities
    \textsuperscript{1} University of Louisiana at Lafayette \quad\quad\quad  \textsuperscript{2} East China Normal University \quad\quad\quad \textsuperscript{3} University of Delaware %
    }
}

\iffalse
\author{\IEEEauthorblockN{Abeda Sultana}
\IEEEauthorblockA{\textit{School of Computing and Informatics \\University of Louisiana at Lafayette \\
Lafayette, LA, USA\\
 abeda.sultana1@louisiana.edu}\\
  }
\and
\IEEEauthorblockN{Nabin Pakka}
\IEEEauthorblockA{\textit{School of Computing and Informatics \\University of Louisiana at Lafayette \\
Lafayette, LA, USA\\
 nabin.pakka1@louisiana.edu}\\
  }
\and
\IEEEauthorblockN{Li Chen}
\IEEEauthorblockA{\textit{School of Computing and Informatics \\University of Louisiana at Lafayette \\
Lafayette, LA, USA\\
li.chen@louisiana.edu\\}
}
\and
\IEEEauthorblockN{Fei Xu}
\IEEEauthorblockA{\textit{Department of Computer Science and Technology \\
East China Normal University\\
China\\
fxu@cs.ecnu.edu.cn\\}
}
\and
\IEEEauthorblockN{Xu Yuan}
\IEEEauthorblockA{\textit{School of Computing and Informatics\\
University of Louisiana at Lafayette\\
Lafayette, LA, USA\\
xu.yuan@louisiana.edu\\}
}
}
\renewcommand{\shortauthors}{Abeda and Pakka, et al.}
\fi
\maketitle

\begin{abstract}
Scheduling deep learning (DL) models to train on powerful clusters with accelerators like GPUs and TPUs, presently falls short, either lacking fine-grained heterogeneity awareness or leaving resources substantially under-utilized. To fill this gap, we propose a novel design of a task-level heterogeneity-aware scheduler,  \textit{Hadar}, based on an optimization framework that can boost resource utilization. \textit{Hadar} leverages the performance traits of DL jobs on a heterogeneous DL cluster, characterizes the task-level performance heterogeneity in the optimization problem, and makes scheduling decisions across both spatial and temporal dimensions.  
%with the objective to reduce the average job completion time of DL jobs. 
It involves the primal-dual framework employing a dual subroutine, to solve the optimization problem and guide the scheduling design. Our trace-driven simulation with representative DL model training workloads demonstrates that \textit{Hadar} accelerates the total time duration by 1.20$\times$ when compared with its state-of-the-art heterogeneity-aware counterpart, Gavel. 
Further, our \textit{Hadar} scheduler is enhanced to \textit{HadarE} by forking each job into multiple copies to let a job train concurrently on heterogeneous GPUs resided on separate available nodes (i.e., machines or servers) for resource utilization enhancement. \textit{HadarE} is evaluated extensively on physical DL clusters for comparison with \textit{Hadar} and Gavel. With substantial enhancement in cluster resource utilization (by 1.45$\times$), \textit{HadarE} exhibits considerable speed-ups in DL model training, reducing the total time duration by 50\% (or 80\%) on an Amazon's AWS (or our lab) cluster, while producing trained DL models with consistently better inference quality than those trained by \textit{Hadar}.
\end{abstract}


% Customized line before the footnote
\renewcommand{\footnoterule}{
    \kern -3pt
    \hrule width 1\columnwidth height 0.5pt
    \kern 2.6pt
}
\renewcommand{\thefootnote}{} % footnote without number
\footnotetext{
\textsuperscript{\scriptsize \(\dagger\)} A preliminary version of this article \cite{hadar} was presented at 38\textsuperscript{th} IEEE Int'l Parallel and Distributed Processing Symposium (IPDPS), May 2024.
\vspace{2pt}

This work was supported in part by the National Science Foundation under Grants: OIA-2019511, OIA-2327452, CNS-2315613, CNS-2348452, in part by the Louisiana Board of Regents under Contract LEQSF(2019-22)-RD-A-21, in part by NSFC under Grant 62372184 and by STC of Shanghai Municipality under Grant 22DZ2229004.}

\section{\textbf{Introduction}}
\label{sec:intro}
\thispagestyle{plain}

Deep learning (DL) applications are ubiquitous nowadays across various domains, including speech recognition, natural language processing \cite{nlp}, super-computing, social media \cite{social_media}, among others. To facilitate the ever-increasing demand for DL training \cite{gu2019tiresias}, large enterprises and cloud providers \cite{AWS,Azure,Google_cloud} have constructed powerful DL clusters, which usually incorporate specialized accelerators, such as GPUs, TPUs, and FPGAs, to accelerate DL model training with intricate architectures. The wide adoption of DL models calls for training multiple of them concurrently on such DL clusters with expensive resources. Efficiently scheduling multiple DL training jobs is thus required to yield high training performance, measured by cluster-wide resource utilization, the total training time duration, the average job completion time, {\em etc.}

To this end, existing efforts have proposed a number of GPU cluster schedulers ({\em e.g.,} \cite{xiao2018gandiva,gu2019tiresias,philly}) for DL model training. 
%including Gandiva \cite{xiao2018gandiva}, Tiresias \cite{gu2019tiresias}, Philly \cite{philly}, and {\em etc.} However, they 
However, those schedulers either lack the awareness of job or under-utilize available resources, leading to undesirably low performance for DL clusters.
%The unique characteristics of deep learning training jobs, elaborated in what follows, make it suboptimal or impractical to employ legacy schedulers in traditional clusters for big data analytics. 
%The distinctive characteristics of deep neural networks elaborated as follows. 
%{\em First,} trained in a distributed manner, as single device is far from sufficient for the enormous computation demands. With the prevailing data parallel training mode, a deep learning job is trained with a large number of epochs, spanning a number of devices for parallel training tasks. {\em Second}, given a wide array of deep learning applications, training jobs are heterogeneous in nature, differing to a great extent in resource demands and performance requirements. {\em Third}, 
It has been observed in \cite{gavel} that DL training jobs show heterogeneous performance behavior across accelerator devices of different types, due to various architectural differences.
For example, a ResNet-50 model achieves a nearly 10$\times$ speedup when trained on an NVIDIA V100 GPU versus a K80 GPU, while an A3C Deep Reinforcement Learning model only exhibits 2$\times$ acceleration. %Moreover, spanning a job across devices of multiple types could benefit scheduling policies designed for various objectives ({\em e.g.}, dollar cost efficiency). 
In light of such observations, Gavel has been proposed \cite{gavel} as a heterogeneity-aware cluster scheduler, which is the first to address the aforementioned performance heterogeneity of DL training jobs across multi-type accelerators in a cluster. 
It utilizes an optimization-based scheduling framework to specifically account for job placement and performance heterogeneity.
%However, it does not study the impact of heterogeneity at task-level granularity for DL training jobs. In other words, the scheduling framework in Gavel \cite{gavel} does not explicitly characterize the performance heterogeneity at finer-grained task level ({\em i.e.}, tasks of different workloads on heterogeneous devices), which is crucial to mitigating the straggler issue and thus the performance of DL training jobs.
However, it does not explicitly characterize performance heterogeneity at a fine-grained task level, with DL workloads scheduled at the job level. If a job requires 4 V100 GPUs, but the cluster has 3 V100 and 3 K80 GPUs available, the job cannot proceed and must wait for the next scheduling round. This limitation highlights the need for a more sophisticated and flexible scheduler that can make the best use of the available cluster resources by accommodating task-level performance heterogeneity.

To bridge this gap, we first introduce a new fine-grained heterogeneity-aware scheduler, named {\em Hadar} \cite{hadar}, for a cluster shared by DL training jobs. 
%focused on improving resource usage and minimizing the overall job completion time of the cluster. 
The essence of {\em Hadar} relies on the problem formulation and optimization framework for task-level resource allocation across both temporal and spatial dimensions, as opposed to the state-of-the-art counterpart, Gavel \cite{gavel}, whose optimization framework only characterizes the spatial resource allocation with just job-level heterogeneity awareness. Trace-driven simulation is adopted to evaluate {\em Hadar}, for comparison with its previous counterparts, including Gavel \cite{gavel}, Tirsias \cite{gu2019tiresias}, and YARN-CS \cite{Azure}. Simulation results demonstrate that {\em Hadar} solidly outperforms in terms of such metrics as resource utilization, total training time duration, and scalability.

Next, {\em Hadar} is further enhanced by forking every training DL job into multiple copies for possible concurrent execution on heterogeneous GPUs resided on different nodes (i.e., machines or servers) of a DL cluster, boosting cluster resource utilization. This way enables a DL training job to run on various types of GPUs at different nodes simultaneously, if available, for enhancing cluster resource utilization, arriving at {\em Hadar Enhancement} (or {\em HadarE} for short). During the course of job training, {\em HadarE} lets any unfinished job be executed on as many available GPU-equipped nodes as possible due to its forked copies in existence, unlike {\em Hadar} which schedules one job to run merely on a single GPU-equipped node even when another GPU-equipped node is idle. Our {\em HadarE} enables to start its scheduling immediately and effectively without undergoing job profiling \textit{ a priori}, common to earlier  schedulers.  For evaluating {\em HadarE} in comparison with {\em Hadar} and Gavel, we conduct real-world experiments on physical DL clusters leased from the AWS Cloud and at our research lab.

Our extensive experiments on an AWS (or our lab testbed) cluster exhibit that \textit{Hadar} achieves a 20\% (or 21\%) increase in cluster resource  utilization and the speedup of 1.17$\times$ (or 1.16$\times$) in terms of the total time duration across seven different workload mixes with 1 to 12 jobs. \textit{HadarE} enjoys the further performance gains of 30\% (or 34\%) in cluster resource utilization, 90\% (or 124\%) in the mean job completion time, and more than 50\% (or 80\%) in the total time duration on an AWS (or our testbed) cluster. Besides its training acceleration, \textit{HadarE} is demonstrated also to train DL models with better inference quality than \textit{Hadar}. Overall, the main contributions of this work are as follows

\vspace{-0.18cm}
\begin{itemize}
\item An efficient scheduler for DL training jobs in GPU is proposed to address the performance heterogeneity of multi-type accelerators at task-level granularity, arriving at {\em Hadar}.

\item An optimization algorithm is developed following the primal-dual framework which employs a dual subroutine to analyze and tackle the scheduling problem on multiple heterogeneous cluster nodes. 

\item We prove the polynomial runtime complexity of our algorithm and also perform a detailed analysis to provide a long-term performance guarantee that approximates the optimal solutions within proven constant bounds.
%It exploits dynamic programming structure to find the best scheduling 

\item Each model training job is forked into multiple copies for possible concurrent executions on separate heterogeneous DL cluster nodes, to further enhance cluster resource utilization, realizing {\em HadarE} ({\em Hadar Enhancement}).
 
\item Extensive experiments are conducted on two physical DL clusters: one leased from the AWS Cloud and the other available at our research lab, with results consistently demonstrating the solid advantages of {\em HadarE} over \textit{Hadar} and Gavel.

\end{itemize}

%TODO: revise this paragraph. the related work in introduction should be very focused and related with our motivation and contribution. The next paragraph should be our difference, observation and motivation. Try your best to make it consistent with the next paragraph, and the logic flow should be smooth. If necessary, you could only talk about a few important works that are mostly related.

%Different with these efforts ... to be continue... talk about our motivation, observation on job epoch progress rate, how they can help improve schedueling. focus on the insight, that motivates our design.

\section{\textbf{Related Work and Motivation}}
\label{sec:related}
With the prevailing data parallel training model, a deep learning (DL) job is trained typically for a large number of epochs and on multiple devices (i.e., GPUs or other accelerators), to process voluminous input data in an iterative manner. The model training job is conducted on the machines of DL cluster, often employing a stochastic gradient descent (SGD) mechanism to keep improving the model's learnable parameters. The training data is fragmented into chunks (i.e., mini-batches) for training acceleration under SGD. A complete pass through the whole training data, with one chunk at a time, is referred to as an epoch \cite{adam}. A DL training job usually involves many epochs, and it stops after finishing a pre-specified number of epochs (say, 60) or finding the model prediction loss stabilized for a given number of consecutive epochs, known as early stopping.

To accommodate multiple DL training jobs, traditional CPU-based cluster schedulers fall short, due to the lack of consideration on the unique characteristics of distributed DNN training. Recent production-scale workload analyses \cite{mlaas,sensetime} %({\em e.g.}, a two-month trace from a production cluster with over 6000 GPUs in Alibaba \cite{mlaas})%
confirm such characteristics in the temporal (job runtime) and spatial (resource request) patterns exists in a DL cluster.
In addition, the heavy-tailed nature of resource requests, along with the wide range of queuing delay and job runtime, has increasingly drawn research attention to DL cluster scheduling (exemplified by the pursuits at \cite{xiao2018gandiva, gu2019tiresias, zhao2022multi}).
Those pursuits focus on designing customized cluster schedulers for DNN training jobs to enhance job performance and resource utilization, but they fail to effectively address the adverse impact of resource heterogeneity.
%, which is a key focus of our proposed scheduler.

%\textbf{Resource Heterogeneity:}
%\textbf{Heterogeneity Aware Deep Learning Cluster manager:}
Some recent schedulers \cite{chaudhary2020balancing,gavel,yang2023hydra} have paid attention to device and model workload heterogeneity to some extent, but not at a fine granularity. Gandiva$_{fair}$ \cite{chaudhary2020balancing} focuses merely on fairness among jobs while Hydra \cite{yang2023hydra} aims at meeting deadlines. Gavel \cite{gavel} accounts for the heterogeneity of both DNN training workloads and hardware devices when allocating resources among jobs. It presents a general optimization framework to characterize a number of scheduling policies. Lately, studies have treated both CPUs and accelerators as prime factors upon assigning workloads in a DL cluster \cite{synergy} \cite{datacenter2024survey}. A recent study, called shockwave \cite{shockwave}, predicts future utility amid dynamic adaption. Being model-agnostic, PPS \cite{pps} treats the DL training job as a black box and predicts future usage considering only job statictics. 

Unlike all prior schedulers, our {\em Hadar} (or {\em HadarE}) is aware of task-level (or job-level) performance heterogeneity. \textit{Hadar} and \textit{HadarE} allocate resources at finer granularity across an additional temporal dimension, yielding marked overall performance improvement.

% In sharp contrast, in this paper, we design a new cluster scheduler for DNN training jobs, effectively incorporating heterogeneity awareness in our online resource scheduling at the task level of finer granularity, across both temporal and spatial dimensions.  Compared to Gavel \cite{gavel}, which is the closest counterpart of our work among the state-of-the-art, our scheduler is aware of task-level performance heterogeneity and allocates resources at finer granularity across an additional temporal dimension, yielding marked overall performance improvement. Moreover, similar to Gavel \cite{gavel}, our optimization framework underlying the scheduler is also able to express other scheduling policies. We next present a toy example to highlight our insights, further strengthening our motivation.

\subsection{\textbf{Motivational Example}}
A cluster with two V100, three P100, and one K80 GPUs is considered. Three jobs arrive at the beginning to be scheduled. Job 1 (J1) requests 3 GPUs and requires 80 epochs to complete its training. Job 2 (J2) requests 2 GPUs and has a total of 30 epochs. Job 3 (J3) requires 2 GPUs for 50 epochs.
The training speedups of those jobs on GPUs of different types, expressed as a matrix $X$, and the optimal allocation matrix $Y^{Gavel}$, under the assumption that the cluster has sufficient capacity, are expressed as 

\begin{minipage}{0.45\linewidth}
\centering
\scalebox{0.70}{$
X = \bordermatrix{~  & V100 & P100 & K80 \cr
              J1 & 40 & 20 & 30\cr
              J2 & 5 & 15 & 5\cr
              J3 & 10 & 2 & 20\cr
              }
$}
\end{minipage}%
\hfill
\begin{minipage}{0.45\linewidth}
\centering
\scalebox{0.70}{$
Y^{Gavel} = \bordermatrix{~  & V100 & P100 & K80 \cr
              J1 & 0.6 & 0.4 & 0.0\cr
              J2 & 0.2 & 0.6 & 0.2\cr
              J3 & 0.2 & 0 & 0.8\cr
              }
$}
\end{minipage}
\vspace{5pt}

\begin{figure}[t]
    \centering
    \begin{minipage}[b]{0.24\textwidth}
        \centering
        \includegraphics[width=0.94\textwidth]{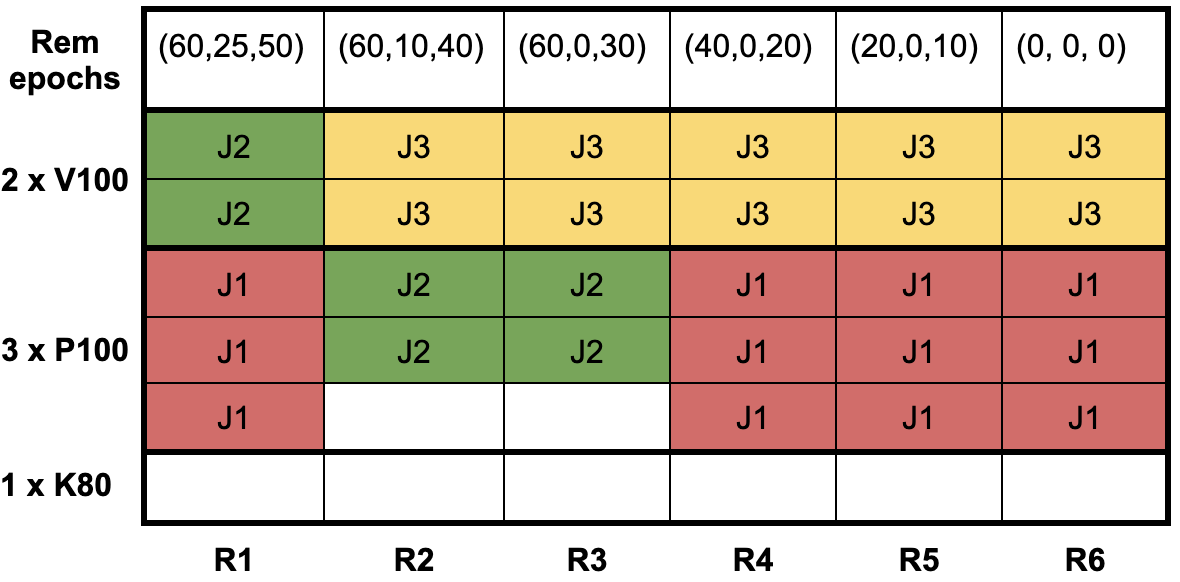}
        \caption*{(a) Round-based simulation of Gavel}
        \vspace{15pt}
        \label{fig:gavel-illustration}
    \end{minipage}
    \begin{minipage}[b]{0.24\textwidth}
        \centering
        \includegraphics[width=0.85\textwidth]{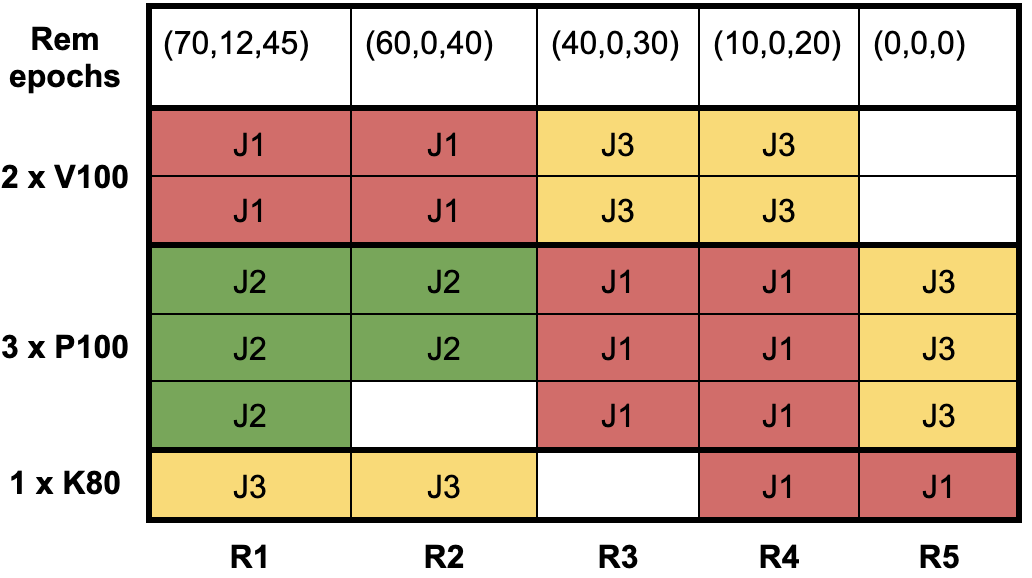}
        \caption*{(b) Round-based simulation of {\em Hadar}}
        \vspace{15pt}
        \label{fig:hadar-illustration}
    \end{minipage}
\caption{Comparative simulation results of scheduling three jobs in a cluster with 2$\times$ V100, 3$\times$ P100, and 1$\times$ K80 GPUs under Gavel \cite{gavel} and {\em Hadar} \cite{hadar}.}
\label{fig:simulation-illustration}

\end{figure}

Each element of this matrix represents the proportion of time that a job should run on a specific type of devices. To achieve a near-optimal allocation, Gavel uses a priority matrix to schedule jobs on GPUs. The priority of a specific job on a particular type of GPUs is defined as the corresponding element of $Y^{Gavel}$ divided by the number of rounds received ({\em i.e.}, resource allocation received).
Fig. \ref{fig:simulation-illustration}(a) illustrates the scheduling outcome over 6 rounds according to Gavel, where the first row represents the number of remaining epochs (i.e., Rem epochs) for each of the three jobs at a particular round. For example, in round 1 (R1 in the figure), J1, J2, and J3 have 60, 25, and 50 epochs to complete, respectively, represented by 60, 25, 50 in Fig. \ref{fig:simulation-illustration}(a). 

Within each round, Gavel schedules all tasks of a job on the same type of GPUs. In contrast, we exploit the flexibility of task-level allocation, to maximize the overall performance of the cluster. As shown in Fig. \ref{fig:simulation-illustration}(b), \textit{Hadar} strategically assigns the tasks of job $J1$ to all GPUs (i.e., two $V100$ GPUs, three $P100$ GPUs, and one $K80$ GPU), tasks of job $J2$ to three $P100$ GPUs, and tasks of $J3$ to all GPUs during the whole process. In comparison, Gavel's policy adopts the homogeneous allocation of tasks strictly, causing jobs $J1$, $J2$ and $J3$ to achieve lower cluster resource utilization (CRU) in the long run. For example, the CRU values of \textit{Hadar} (or Gavel) in the first two rounds approach 100\% (or 83\%) and 83\% (or 67\%), respectively. Over the entire process of scheduling the three jobs, \textit{Hadar} achieves its CRU of some 87\%, versus $\sim$78\% under Gavel, besides shortening the total training time by one round, as depicted in Fig.~\ref{fig:simulation-illustration}.

\section{\textbf{Design of \textit{Hadar}}}
\renewcommand{\arraystretch}{1.25}
\begin{table}[t]
\centering
\hfill
\vspace{5pt}
\caption{Notation}\label{tab:notation} 
\vspace{20pt}
\begin{tabular}{|l|l|}
\hline
$J$           & \# of jobs  \\ \hline
$R$           & \# of GPU types  \\ \hline
%$H$           & \# of machines   \\ \hline
$a_j$       & arrival time of job $j$   \\ \hline
$f_j$       & finish time of job $j$ \\ \hline
$W_j$       & \# of GPUs requested by job $j$  \\ \hline
$E_j$       & \# of total training epochs specified by job $j$    \\ \hline
$N_j$       & \# of data chunks (iterations) per epoch in job $j$ \\\hline
$c_h^r$     & \# of type-$r$ GPUs on machine $h$   \\ \hline
$X_j^r$     & \# of training iterations per {\em sec} for job $j$ on type-$r$ GPU  \\ \hline
$w_{jh}^r(t)$ & \begin{tabular}[c]{@{}l@{}}\# of type-$r$ GPUs on machine $h$ allocated to job $j$ \\at time $t$     \end{tabular} \\ \hline
%$y_{js}$ &  binary decision on job $j$'s admission under schedule $s$\\\hline
$\mathcal{U}_j(.)$ & utility of job $j$ \\ \hline
\end{tabular}
\end{table}
\renewcommand{\arraystretch}{1}

Our overall objective is to design an effective scheduler for distributed DL training jobs with the awareness of resource heterogeneity and best performance. In this section, we present the theoretical foundation of our scheduler design.

\subsection{\textbf{System Model and Problem Formulation}}
Consider a cluster of machines equipped with different accelerator devices. A machine $h$ has a capacity of $c_h^r$ for type-$r$ device. In a slotted time spectrum ($1,2,\cdots,T$), a DL job $j$ arrives at time $a_j\in[T]$, requesting a number of worker devices $W_j$ for model training. 
The device heterogeneity impacts the job training throughput, denoted by $X_j^r$ which represents the number of iterations per second by job $j$ on type-$r$ accelerator. $E_jN_j$  denotes the total number of iterations to complete job $j$, where $E_j$ refers to the total number of epochs and $N_j$ is the total number of data chunks to be processed in each epoch of job $j$.

Upon arrival, the job joins a global queue managed by the scheduler, waiting to be assigned to available machine(s) for execution in subsequent time slots. 
The scheduler makes scheduling decisions, %\textit{i.e.}, which job(s) to schedule, together with the machine and GPU assignment, at the beginning of each time slot. The decision variables, 
$w_{jh}^{r}(t)$, representing the number of type-$r$ devices at machine $h$ assigned to job $j$ in time slot $t$. Let $f_j$ denote the finish time of job $j$, and thus the job completion time can be expressed as $f_j-a_j$.
We start with finding the optimal resource allocation and scheduling to maximize the overall utility across all jobs. The utility of a job $j$ could be characterized by a general non-negative function $\mathcal{U}_j(\cdot)$ which is non-increasing with its completion time. The effective throughput \cite{gavel}, defined as the averaged number of iterations completed per second over the lifetime, can be a special case of the job utility, expressed as $E_jN_j$ divided by $j$'s completion time. Given these notations, we can formulate the following optimization problem, P1:
\vspace{-5pt}
\begin{subequations}\label{opt}
\begin{align}
\max & \quad \textstyle{  	\sum_{j}\mathcal{U}_j(f_j-a_j) } \tag{\ref{opt}}\\
	s.t.  & \quad \textstyle{  	\sum_t x_j(t) \sum_r \sum_h w_{jh}^r(t) L \geq E_jN_j, \ \ \forall j } \label{cons:iterations} \\
	%& & \textstyle{  x_j(t) = min_{r| w_{jh}^r(t)>0}\{X_j^r\}, \ \ \forall j, \  \forall t } \label{cons:throughput}\\
	& \quad \textstyle{  x_j(t) = \min\{X_j^r|\sum_h w_{jh}^r(t)>0\}, \ \ \forall j, \  \forall t } \label{cons:throughput}\\
%	 & & \textstyle{  	0 \leq x_j(t) \leq X_j^r, \ \ \forall  r|\sum_h w_{jh}^r(t)>0 } \\
    & \quad \textstyle{  	f_j = \max\{t\in[T]|\sum_{h}\sum_r w_{jh}^r(t) > 0\}, \ \ \forall j } \label{cons:finish}\\
 %   & & \textstyle{  	f_j \geq t, \ \ \forall  \{t|\sum_h \sum_r w_{jh}^r(t)>0\}, \forall j} \label{eqn:non-negative} \\
 	& \quad \textstyle{  	0 \leq \sum_{j} w_{jh}^r(t) \leq c_h^r,\ \ \forall h, \  \forall r, \  \forall t } \label{cons:capacity}\\
	& \quad\textstyle{  	\sum_{h}\sum_{r} w_{jh}^r(t)\in \{0,W_j\}, \ \ \forall j, \ \forall  t\geq a_j  } \nonumber \\
%	& & \textstyle{  	0 \leq \sum_{h}\sum_{r} w_{jh}^r(t)\leq W_j, \forall j, \forall  t\geq a_j } \\
& \quad \textstyle{  	w_{jh}^r(t)=0, \ \  \forall j, \ \forall h, \ \forall  t< a_j }. \label{cons:alloc}
\end{align}
\end{subequations}
%\end{eqnarray}
Constraints (\ref{cons:iterations}) and (\ref{cons:throughput}) regulate that the total number of iterations accomplished across time is no smaller than $E_jN_j$ to complete job $j$. Specifically, $L$ is the length of a time slot, $x_j(t)$ expresses the bottleneck throughput across tasks, {\em i.e.}, the number of iterations per second at the slowest device, due to the parameter synchronization barrier. Constraint (\ref{cons:finish}) by definition, represents the last time slot when a job receives non-zero allocation to run.
Constraint (\ref{cons:capacity}) indicates resource capacity limits at each machine, while (\ref{cons:alloc}) regulates resource requirements for each job, {\em i.e.}, the All-or-Nothing property (Gang scheduling), following the conventional practice \cite{mlaas}. %\cite{karatza2009performance}). 
A brief notation summary is presented in Table~\ref{tab:notation}.

% {\bf Expressing other scheduling policies.} Note that our optimization-based scheduling framework can express other scheduling objectives. For example, minimizing the average job completion time is denoted as $\min \sum_{j}(f_j-a_j)/J $, minimizing the makespan is represented as $\min \max_{j}f_j$), and achieving fairness across users can be $\min \max_{j}(f_j-a_j)/(f_j^{isolated}-a_j)$, considering the finish-time fairness metric \cite{mahajan2020themis}, where $f_j^{isolated}$ is the job finish time when using $1/J$
% of the cluster. 

\subsection{\textbf{Problem Solving based on Primal-Dual}}
\label{problem-primal-dual}
The optimization problem P\ref{opt} is difficult to solve since it involves integer variables and non-conventional constraints (\ref{cons:throughput}), (\ref{cons:finish}).
\iffalse
We will investigate how to equivalently transform or relax the problem to eliminate the complexities of non-linear constraints, deriving insights to guide us to design practical algorithms and conduct theoretical analysis. 

A few scheduling objectives are to be studied, such as minimizing the average job completion time ($\min \sum_{j}(f_j-a_j)/J $), minimizing the makespan ($\min \max_{j}f_j$), achieving fairness across users (for example, $\min \max_{j}(f_j-a_j)/(f_j^{isolated}-a_j)$ considering the finish-time fairness metric \cite{mahajan2020themis}, where $f_j^{isolated}$ is the job finish time when using $1/J$
of the cluster), {\em etc.}

We will consider another problem variation by changing constraint (\ref{cons:alloc}) to be $\sum_{h}\sum_{r} w_{jh}^r(t)\in [0,W_j]$. This allows job elasticity across scheduling rounds, {\em i.e.}, the degree of parallelism in model training can be flexibly adjusted which could be optionally supported by the machine learning framework \cite{xiao2020antman}. %to be one of our subsequent tasks.
Eventually, we hope to design a pocket of efficient online algorithms for a variety of operation goals, without assuming any knowledge of future job arrivals, and analyze their competitive ratios. \\
\fi
To address these challenges, we first reformulate Problem P\ref{opt} into the following integer linear program (ILP).
Suppose $\mathbb{S}_j$ is the set of feasible schedule for job $j$ which corresponds to the set of decisions ($w_{jh}^r(t), \forall h\in[H], j\in[J], t\in[T]$). It satisfies constraints (\ref{cons:iterations}), (\ref{cons:throughput}), and (\ref{cons:alloc}). Due to the combinatorial nature of these constraints, there is an exponential number of feasible schedules for each job. For a schedule $s\in \mathbb{S}_j$, the decision variable in the ILP is a binary variable $y_{js}$ which indicates whether the job is admitted to the cluster under schedule $s$.  With schedule $s$, job $j$'s finish time is denoted as $f_{js}$, and its allocation $w_{jh}^{rs}(t)$ represents the number of type-$r$ workers in server $h$ at time $t$. Thus, P\ref{opt} can be reformulated to P\ref{opt2} as follows:
\begin{subequations}\label{opt2}
\begin{align}
\max & \quad	\textstyle{  \sum_{j } \sum_{s } y_{js}\mathcal{U}_{j}(f_{js}-a_{j})} \tag{\ref{opt2}}\\
s.t. & \quad	\textstyle{  \sum_{j} \sum_{s:t\in s, h \in (t,s)}  w_{jh}^{rs}(t) y_{js} \leq c_h^r, \forall h ,\forall r ,\forall t} \label{p2a}\\
& \quad	\textstyle{ \sum_{s } y_{js} \leq 1, \forall j} \label{p2b}\\
& \quad	\textstyle{ y_{js} \in \{0,1\}, \forall j,\forall s }.\label{p2c}
\end{align}
\end{subequations}
We use $t \in s, h \in (t, s)$ to indicate that schedule
$s$ uses server $h$ to deploy a worker for job $j$ in time $t$. Eq.~(\ref{opt2}) and constraint (\ref{p2a}) are equivalent to Eq.~(\ref{opt}) and constraint (\ref{cons:capacity}), respectively. Constraints (\ref{p2b})-(\ref{p2c}) are equivalent to constraints (\ref{cons:iterations})-(\ref{cons:finish}) and (\ref{cons:alloc}). We can easily check that P\ref{opt} and P\ref{opt2} are equivalent, since a feasible solution to one has a corresponding feasible solution to the other, with the same objective values.

After sidestepping non-conventional constraints, we next 
%The number of variables in (\ref{opt2}), $y_{js}$'s, is potentially exponential, we will 
solve Problem P\ref{opt2} based on the primal-dual framework \cite{TCS}, by relaxing its integer constraints (\ref{p2c}) and formulating its dual problem designated as P3 below:
\begin{subequations}\label{p3}
\begin{align}
\min & \quad \textstyle{ \sum_j  \mu_j + \sum_{t}\sum_{h }\sum_{r } k_h^r(t) c_h^r(t)} \tag{\ref{p3}}\\
s.t. & \quad
\textstyle{ \mu_j \geq \mathcal{U}_{j}(f_{js}-a_{j})} -  \nonumber \\
 & \quad \textstyle{ \sum_{t \in s}\sum_{h \in (t,s)}\sum_{r} k_h^r(t) w_{jh}^{rs}(t), \forall j,\forall s} \label{p31}  \\
& \quad \textstyle{ k_h^r(t)\geq 0, \forall  h , \forall  r ,\forall  j ,\forall t, \quad \quad \mu_j \geq 0, \forall j}. \nonumber
\end{align}
\end{subequations}
In this problem, $k_h^r(t)$ and $\mu_j$ are the dual variables associated with constraints (\ref{p2a}) and (\ref{p2b}).
$k_h^r(t)$ can be interpreted as the unit cost for type-$r$ accelerators on server $h$ at time $t$.
%$\sum_{h \in (t,s)}\sum_{r \in [R]} k_h^r(t) w_{jh}^{rs}(t)$ is thus the total resource cost of all tasks of job $j$ with schedule $s$ in time $t$. 
Thus, the right-hand side of (\ref{p31}) is the job utility minus the overall resource cost for job $j$ with schedule $s$ at time $t$, which indicates the payoff of the job.
Let $\phi_j(s)$ denote this term, {\em i.e.}, $\phi_j(s) = \mathcal{U}_{j}(f_{js}-a_{j}) - \sum_{t \in s}\sum_{h \in (t,s)}\sum_{r} k_h^r(t) w_{jh}^{rs}(t)$.
To minimize the dual objective, $\mu_j^*$ should be expressed as $\mu_j^* = \max\{0, \max_{s \in \mathbb{S}_j} \phi_j(s) \}$, based on its constraints. 
The corresponding best schedule $s^*$ can be written as
\begin{align}\label{p4}
& s^* = argmax_{s \in \mathbb{S}_j} \phi_j(s). 
\end{align}

To solve Eq.~(\ref{p4}),  we design an efficient subroutine to be elaborated later (Algorithm \ref{algo2}).
With respect to $k_h^r(t)$, based on its resource price interpretation, we hope to compute its value to ensure that a high-utility job gets a positive payoff (if the resource demand can be satisfied) and a job with a low utility or without available resources gets a non-positive payoff. 
Let $\gamma_h^r(t)$ denote the number of type-$r$ accelerators allocated on server $h$ at time slot $t$. The dual price resource is designed to be dynamically updated using the following price function:

{ \small
\begin{align}
& \quad k_h^r(\gamma_h^r(t)) = U_{min}^r(\frac{U_{max}^r}{U_{min}^r})^{\frac{\gamma_h^r(t)}{c_h^r }},\label{eqn5}\\
with & \quad U_{max}^r = \max_j \frac{\mathcal{U}_j(t^{min}_j -a_j)}{ w_j^r}, ~~~ \forall r \label{defU}\\
& \quad U_{min}^r =  \frac{1}{4\eta}\min_{j } \frac{\mathcal{U}_j(T -a_j)}{ t_j^{max}\sum_{r\in[R]} w_j^r}, ~~~ \forall r \label{defL}\\
& \quad t^{min}_j = \frac{N_jE_j}{M_j \max_{r} (X^r_j)}, \quad t^{max}_j = \frac{N_jE_j}{M_j \min_{r} (X^r_j)},\notag 
\end{align}
}
\noindent where $U_{max}^r$ and $U_{min}^r$ imply the maximum and the minimum per-unit-resource job utility values for type-$r$ accelerators to execute tasks among all jobs. $\mathcal{U}_j(T -a_j)$ is the smallest utility that job $j$
may achieve, when it ends at $T$. $\eta$ is the scaling factor %satisfying $\frac{1}{\eta} \leq \frac{t_j^{max }\sum_{r\in[R]}w_j^r}{ \sum_{h \in [H]}\sum_{r \in [R]}c_h^r}, \forall j \in [J]$,  
to bound the initial value of the dual objective.

The intuition is stated as follows. 
The price starts to be low enough to accept the incoming job: when $\gamma_h^r = 0$, we have $k_h^r(t) = U_{min}^r$, lowest to admit any job.
% thus any job can be admitted as $L$ is the lowest utility. 
The price increases exponentially with the growing amount of allocated accelerators, so as to filter out low-utility jobs.
When a server is out of free resources, $\gamma_h^r(t)$ = $c_h^r$, reaching the price $k_h^r(t) = U_{max}^r$, high enough to block other jobs from getting these resources.  % and no more job requiring this type of accelerators can be admitted. 
Such a price function is crucial to guarantee a good competitive ratio for our effective algorithm, to be presented in the next section. $U_{max}^r$ and $U_{min}^r$ are calculated based on the cluster's workload in the algorithm.

\subsection{\textbf{Algorithm Design}}
Based on the resource price and job payoff interpretations, we next present our algorithm (Algorithm \ref{algo1}), which generates optimal scheduling decisions for the jobs in the queue in each 
%an online algorithm is designed where the best schedule of job $j$ from the queue is computed in each 
round-based scheduling event (\texttt{Line 5}).
Specifically, a greedy algorithm and a dynamic programming approach are presented in Algorithm~\ref{algo2},
to calculate $s^*$ in Eq.~(\ref{p4}) by solving the following equivalent form: 
\begin{align}
    \max & \quad \textstyle{ \mathcal{U}_{j}(f_{j}-a_{j}) -\sum_{t }\sum_{h }\sum_{r } k_h^r(t) w_{jh}^r(t) } \label{p8} \\
     s.t. & \quad \gamma_h^r(t) + w_{jh}^r(t) \leq c_h^r,~~~\forall j~in~queue , \forall r ,  \forall h, \forall t  \notag\\
    & \quad \text{Constraints } (\ref{cons:iterations} - \ref{cons:alloc}). \notag
\end{align}

If we fix $f_j$, the optimization objective can be further transformed to $\min \sum_{h }\sum_{r } k_h^r(t) w_{jh}^r(t)$, which can be interpreted as minimizing a cost function at each round.

In Algorithm \ref{algo2}, waiting jobs in the current round are in queue $Q$. According to the recursive dynamic programming solution in each state, there are two possible choices for a certain job, either calculating the cost and allocation by selecting the job for scheduling or proceeding without selecting the job in \texttt{Lines 14-15}. The set of jobs and the allocations with minimum cost is returned from the DP function call \texttt{Lines 16-21}. Note that we always save the result if $cost_Q$ and $cost_{Q/j}$ are compared for different subsets of jobs to avoid recomputing the same subproblem in later recursive function call. The \texttt{FIND\_ALLOC} function selects the best possible allocation within the current state of the server (\texttt{$srvr$}). Initially, the server's state is sorted according to the descending order of throughput (iterations per second) on each GPU type for the job (\texttt{Line 23}). The algorithm produces the allocations on different settings - by consolidating tasks of the job in the minimum possible server (\texttt{Line 24}) and allocating the tasks of the job in different servers (\texttt{Line 25}). The costs are calculated using the cost function aforementioned. For non-consolidated setting, communication cost (the cost of bandwidth utilization while communicating among different servers) is also added (\texttt{Lines 26-27}). The allocation with the minimum cost is calculated, and $\mu_j$ is calculated to determine allocation's feasibility (\texttt{Lines 28-32}). According to the selected allocation, allocated resource $\gamma_h^{rc}(t)$, price function $k_h^{rc}(t)$, and the server state are updated (\texttt{Lines 10-12}).
\subsection{\textbf{Theoretical Analysis}}
\label{theory}
\begin{theorem}[Runtime Complexity]
\label{theo-runtime}
%In an execution round with a set of jobs, Algorithm \ref{algo2} runs in polynomial time to make scheduling decisions. 
Algorithm \ref{algo2} can make scheduling decisions in polynomial time for a set of jobs in an execution round.
\end{theorem}
\vspace{-10pt}

\begin{algorithm}[t]
\caption{\textit{Hadar} Scheduling}\label{algo1}
   \begin{algorithmic}[1]
        \Require{$c_h^r,\forall h \in [H], r \in [R]$}
        %\Ensure{$ w_{jh}^r(t)\forall j\in [J], t \in [T], h \in [H]$}
       \State \textbf{Initialize:} $ w_{jh}^r(t)(t) = 0, \gamma_h^r(t)=0, k_h^r(t) =  k_h^r(0) \forall j\in [J], t \in [T], h \in [H] $ 
        % \While{ $i = 1,2,...$}
        %     \While{ $t < \tau_i$}
        %         \State $J_i = J_i \cup {j}$
        %     \EndWhile
        \While{true}
        \State Upon the arrival of each job, admit it to the queue $Q$
       
        \State In each round $t$:
            \State $\{{Q_s, c^r_h, \{w_{jh}^r(t)(t)\}}\}$ =
            $ DP\_allocation (0,Q, {c_h^r},null,\gamma_h^r(t), k_h^r(t))$
            % $Hadar_{dual}(J_i, \tau_i, {c_h^r})$
            \For{job $j \in [Q_s]$}
                \State Run job $j$ until round $t+1$ according to $( \{w_{jh}^r(t)\})$
                
            \EndFor
            \State If $j$ is complete, remove it from $Q$
            
        \EndWhile
   \end{algorithmic}
\end{algorithm}

\noindent \textit{Proof.} The function \texttt{FIND\_ALLOC}(\(job, srvr\)) has a time complexity of \(\mathcal{O}(R(H\log H))\) to sort the servers based on job throughput on GPUs of different types. This sorting calculation is only done once during the lifespan of a job in the system. For calculating allocation in both consolidated (\(all\_alloc_{packed}\)) and non-consolidated (\(all\_alloc_{!packed}\)) settings, all servers need to be iterated for each GPU type, resulting in a complexity of \(\mathcal{O}(HR)\). In our dynamic programming (DP) algorithm, we have two states: job ID and the current server state. We need to calculate \(n(Q)HR\) combinations or function calls, with a time complexity of \(\mathcal{O}(HR)\) for each call. It should be noted that we pre-calculate and save \(cost(DP\_allocation(jobs, srvr))\) for all \(j \in Q\). Therefore, the time complexity of the DP is \(\mathcal{O}(n(Q)(HR)^2 + R(H \log H))\).

\vspace{-3pt}

\begin{theorem}[Competitive Ratio]
\textit{Hadar} is $2\alpha$ competitive, where $\alpha = \max_{r \in [R]}(1, ln \frac{U_{max}^r}{U_{min}^r})$ and $U_{max}^r$, $U_{min}^r$ are defined in Eqs.~(\ref{defU}), (\ref{defL}).
\end{theorem}
\vspace{-8pt}
\noindent \textit{Proof.} We define $OPT$ as the optimal objective value of Problem P\ref{opt}. $P_j$ and $D_j$ represent the objective values of the primal problem P\ref{opt2} and of the dual problem P\ref{p3}, respectively, returned by Algorithm \ref{algo1} after deciding the schedule of job $j$. The initial values of Eqs.~(\ref{opt2}) and (\ref{p3}) are denoted by $P_0$ and $D_0$. Specifically, $P_0 = 0$ and $D_0 = \sum_t \sum_h \sum_r k_h^r(0)c_h^r(0)$. Finally, $P_f$ and $D_f$ represent the final primal and dual objective values returned by Algorithm \ref{algo1}. The theorem is proved based on the following definitions and lemmas taken from \cite{oaiss}, to ensure that solutions so derived are bounded within $2\alpha$ from actual optimality.

\vspace{-10pt}
\begin{lemma}
\label{lemma1}
If there exists a constant $\alpha\geq 1$ such that $P_j - P_{j-1} \geq \frac{1}{\alpha}(D_j - D_{j-1})$ for all jobs $j \in [J]$, and if $P_0 = 0$
and $D_0 \leq \frac{1}{2}OPT$, then Algorithm \ref{algo1} is $2\alpha$-competitive in total job utility.
\end{lemma}
\vspace{-20pt}

\begin{definition}
\label{def1}
The allocation-cost relationship for Algorithm \ref{algo1} with $\alpha \geq 1$ is: 
$k_h^{r,j-1}(t)(\gamma_h^{r,j}(t) - \gamma_h^{r, j-1}(t)) \geq \frac{c_h^r}{\alpha} (k_h^{r,j}(t) - k_h^{r, j-1}(t))$. 
\end{definition}
\vspace{-20pt}
\begin{lemma}
\label{lemma2}
If the allocation-cost relationship holds for $\alpha \geq 1$, then Algorithm \ref{algo1} ensures $P_j - P_{j-1} \geq \frac{1}{\alpha} (D_j - D_{j-1}),\forall j$.
\end{lemma}
\vspace{-20pt}
\begin{definition}
\label{def2}
 The differential allocation-cost relationship for Algorithm \ref{algo1} with $\alpha_h^r \geq 1$ is: $k_h^r(t)d\gamma_h^r(t) \geq \frac{c_h^r}{\alpha_h^r} dk_h^r(t), \forall t ,h,r$.
\end{definition}
\begin{algorithm}[t]
  \caption{\texttt{DP\_allocation}($idx, Q, srvr, \{w_{jh}^r(t)\},\gamma_h^r(t),
  \\k_h^r(t), \forall h, \forall r$) }\label{algo2}
 
   \begin{algorithmic}[1]
        % \State $bool$ = $is\_Server\_Full(srvr)$
        %\State $bool_2$ = $Q[idx].is_{scheduled}$
        \If{$(index >= Q.length() )|| is\_Server\_Full(srvr)$}
            \State  \Return $Q, \{w_{jh}^r(t)\}, srvr$
        \EndIf
        \State  $job = Q[idx]$
        \State $\{w\_prev_{jh}^{r}\}\gets \{w_{jh}^r(t)\}$
        \State  $\{w\_job_{jh}^r\}$ = $\textbf{FIND\_ALLOC}(job, srvr)$ 
        \If{$\{w\_job_{jh}^r\}$ = $null$}
            \State \Return $Q, \{w_{jh}^r(t)\}, srvr$
         \EndIf
        \State $\gamma_h^{rc}(t)$ = $\gamma_h^r(t) + w\_job_{jh}^r, \forall h, \forall r$ 
        \State $ k_h^{rc}(t) \gets $ Update $ k_h^r(t)  $ according to Eq.~(\ref{eqn5}), $\forall h, \forall r$
       
        \State  $ srvr^{c} \gets$ Update $srvr$  according to $\{w\_job_{jh}^r\}$
         \State  $\{w_{jh}^r(t)\}.append({w\_job_{jh}^r}), \forall h, \forall r$ 
        \State  $(Q, \{w_{jh}^r(t)\},srvr^{c}) = DP\_allocation $
        \Statex  $((idx+1), Q, srvr^{c},  \{w_{jh}^r(t)\},\gamma_h^{rc}(t),k_h^{rc}(t)), \forall h, \forall r$
        \State  $(Q, \{w_{jh}^{cr}(t)\},srvr) = DP\_allocation $
        \Statex $((idx+1), Q, srvr,\{w\_prev_{jh}^{r}\},\gamma_h^r(t),k_h^r(t)),  \forall h, \forall r$
        \State  $cost_{Q} +=$ $ \sum_h \sum_r k_h^{rc}(t) ~ w_{jh}^r(t) $
        \State $cost_{Q/j} +=$ $ \sum_h \sum_r k_h^{r}(t) ~ w_{jh}^{cr}(t)  $
        \If{ $cost_Q < cost_{Q/j}$}
            \State  \Return $Q, \{w_{jh}^r(t)\}, srvr^c$
        \EndIf
        \State  \Return $Q, \{ w_{jh}^{cr}(t) \}, srvr$

    \Procedure{\textbf{find\_alloc}}{$job, srvr$}:
        \State $srvr \gets$ sort GPU type according to the descending order of $x_j^r,~\forall h \in H$  
        \State      \{$all\_alloc_{packed}\} \gets$find allocations considering consolidated setting
        \State  \{$all\_alloc_{!packed}\} \gets$ find allocations without consideration of consolidated setting

        \State \{$cost_{packed}\} \gets \sum_h \sum_r k_h^r(t)w_{jh}^r(t) , \forall all\_alloc_{packed} $
        
        \State $\{cost_{!packed}\} \gets \sum_h \sum_r k_h^r(t)w_{jh}^r(t) + comm.~cost, \forall all\_alloc_{!packed}$
        
        %  \State $\{cost_{packed}\} \gets$ \texttt{calc\_cost}$(all\_alloc_{packed}$) 
        % \State  $cost_{!packed} \gets$ \texttt{calc\_cost}$(all\_alloc_{!packed} $)  
        
       \State    $alloc \gets$ allocation corresponding to $\min(\{cost_{packed}\}, \{cost_{!packed}\})$ 
      
       \State $\mu_j = \mathcal{U}_{j}(f_{js}-a_{j}) - \min(\{cost_{packed}\}, \{cost_{!packed}\})$
     
       \If {$\mu_j > 0$}
        \State \Return $alloc$
       \EndIf
       \State \Return $null$
      \EndProcedure
    \end{algorithmic} 
\end{algorithm}
\vspace{-20pt}
\begin{lemma}
\label{lemma3}
$\alpha_h^r =ln \frac{U_{max}^r}{U_{min}^r}$ and the price function
defined in Eq.~(\ref{eqn5}) satisfies the differential allocation-cost relationship. 
\end{lemma}
\vspace{-10pt}
Based on Lemma \ref{lemma3}, the marginal cost function employed in Algorithm \ref{algo1} meets the condition of differential allocation-cost relationship with $ \alpha = \max_{r \in R} (1, ln\frac{U_{max}^r}{U_{min}^r}) $. As the resource demand of a job $j$ is expected to be less than the capacity, we can infer that:
$$  d\gamma_h^r(t) = \gamma_h^{r,j}(t) - \gamma_h^{r,j-1}(t)$$
$$  dk_h^r(t)= k_h^{r'}(\gamma_h^r(t))(\gamma_h^{r,j}(t) - \gamma_h^{r,j-1}(t)) = k_h^{r,j}(t) - k_h^{r,j-1}(t).$$

The allocation-cost relationship in Definition \ref{def1} holds for\\ $\alpha = \max_{r } (1, ln \frac{U_{max}^r}{U_{min}^r})$, which is implied by the differential allocation-cost relationship in Definition \ref{def2}. Considering Algorithm \ref{algo1}, we note that $\frac{1}{\eta} \leq \frac{t_j^{max} \sum_{r\in[R]}w_j^r}{\sum_{h }\sum_{r }c_h^r}$, which implies that $\frac{\sum_{h}\sum_{r } c_h^r}{\eta} \leq t_j^{max}\sum_{r } w_j^r$ for all jobs $j$. This minimum resource consumption across all tasks of job $j$ equals: 
{\small
\begin{align}\label{proof-4}
 D_0  & \quad = \sum_{t}\sum_{h} \sum_{r} U_{min}^rc_h^r \\
& \quad =\sum_{t}\sum_{h} \sum_{r} \frac{1}{4\eta} \min_{j, s} \frac{\mathcal{U}_j(f_{js} -a_j)}{t^{max}_j \sum_{r} w_j^r} c_h^r \notag\\
& \quad = \frac{\sum_{t}\sum_{h} \sum_{r}  c_h^r }{4\eta} \min_{j, s} \frac{\mathcal{U}_j(f_{js} -a_j)}{t^{max}_j \sum_{r } w_j^r}\notag \\
& \quad  \leq \frac{1}{4} t_j^{max} \sum_{r } w_j^r \min_{j, s} \frac{\mathcal{U}_j(f_{js} -a_j)}{t^{max}_j \sum_{r } w_j^r}, \forall j. \notag
\end{align} 
}

Selecting $(j,s) = arg \min_{j , s } \mathcal{U}_j(f_{js} -a_j)$ yields: 
{\small
\begin{align}\label{proof-42}
(\ref{proof-4}) & \quad \leq \frac{1}{4} t_j^{max} \sum_{r \in [R]} w_j^r \min_{j, s} \notag \frac{\mathcal{U}_j(f_{js} -a_j)}{t^{max}_j \sum_{r } w_j^r}, \forall j \\ \notag
 & \quad \leq \frac{1}{2}\mathcal{U}_j(f_{js} -a_j) \quad \leq  \frac{1}{2} OPT.
\end{align}
}
The last inequality holds because we assume the offline optimal solution accepts at least one job, which is reasonable in the real-world cluster. Then we have $OPT \geq \min_{j , s }\mathcal{U}_j(f_{js} -a_j)$.
According to Lemmas \ref{lemma1} and \ref{lemma2}, we conclude the proof. \hfill $\qedsymbol$

\subsection{\textbf{Hadar Overview}}
\label{sec:scheduling_policy}

\begin{figure*}
\centering
\includegraphics[width=0.85\textwidth, height=0.15\textwidth]{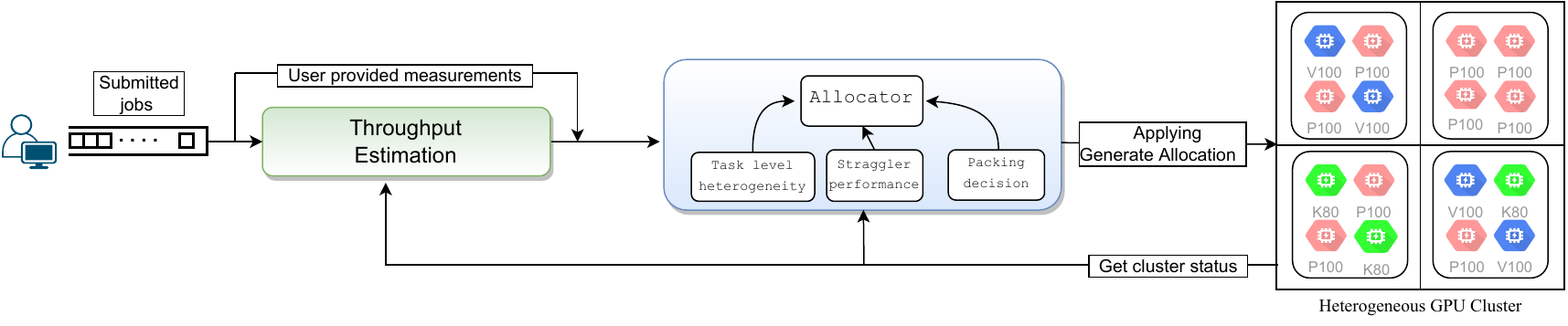}
\caption{The overview of \textit{Hadar}, a fine-grained heterogeneity-aware scheduler for a GPU-based deep learning cluster.}
\label{fig-overview}
\end{figure*}
Guided by the theoretical investigation, our fine-grained heterogeneity-aware scheduler, {\em Hadar}, is illustrated in Fig. \ref{fig-overview}.
Given a set of queued jobs, the scheduler dispatches all jobs onto different types of accelerators on different servers (i.e., machines or cluster nodes) towards maximizing the cluster-wide utility. %The policy needs to compute the number and types of accelerators on different servers for the optimal allocation in a given round. 
Our scheduler takes the job's performance result ({\em i.e.}, iterations per second) on each accelerator type as its input. 
%It can be given either by the user or the throughput estimator of the system.
In particular, the throughput estimator in \textit{Hadar} requires performance measurements for every runnable job on each available accelerator type, and such measurements can be provided either as input data (for trace-driven evaluation in Section ~\ref{sec:trace-driven-eval}) or by an estimation formula (to be detailed in Section ~\ref{sec:physical-cluster-eval}). For a given input, the scheduling algorithm in the allocator calculates the number and types of GPUs assigned to each job on particular machines in a given round. It considers task-level heterogeneity and job packing decisions to maximize overall cluster utility.

\section{\textbf{Trace-Driven Evaluation}}
\label{sec:trace-driven-eval}
Extensive trace-driven simulation is conducted by our discrete-time simulator using a real-world trace \cite{philly} to evaluate \textit{Hadar} for comparison with its counterpart schedulers. Following the setup of the simulation experiments in Gavel \cite{gavel}, our simulated cluster consists of 15 nodes, which house 60 GPUs in total, with 20 GPUs each for V100, P100, and K80.

The workloads are based on a Microsoft trace \cite{philly}, summarized in Table~\ref{tab-models} and elaborated in the next paragraph. For each job (workload) in Table~\ref{tab-models}, we leverage its throughput measurements from Gavel as our scheduling input to simulate the job events such as job arrival, completion, and preemption. The overhead of each checkpoint-restarts is simulated by enforcing a 10-second delay when a job receives a new allocation. According to our evaluation, the duration of a scheduling round impacts the results of evaluation performance metrics, with the duration ranging from 6 minutes to 1.5 minutes to yield the best results, depending on workloads, available resources, and metrics of interest. The results in this section are obtained under the duration of 6 minutes.
 %As the simulator cannot determine job training time with the dynamic cluster environment, it uses actual job completion times. 

\subsection{\textbf{Synthetic Workloads and Datasets}}
In our trace-driven evaluation, we randomly selected 480 jobs from the busiest hour range (hours 3-10) of the Microsoft trace \cite{philly}. The trace includes information such as the requested number of GPUs, submission time, and job duration, while details on model architectures and datasets are not provided. Therefore, we categorized the jobs based on their total GPU required time into four groups: Small (0-1 GPU-hours), Medium (1-10 GPU-hours), Large (10-50 GPU-hours), and XLarge (60-100 GPU-hours). For each training job in the trace, we uniformly sampled the job type from these categories and specified its model and dataset accordingly, as shown in Table~\ref{tab-models}. In our evaluation, all jobs were available at the beginning of the trace.

\subsection{\textbf{Evaluation Results}}
We conducted evaluation for comparing \textit{Hadar} and its state-of-the-art DL cluster scheduler counterparts, Gavel \cite{gavel} and Tiresias \cite{gu2019tiresias}, as well as the default production-level cluster scheduler, Apache YARN's capacity scheduler (YARN-CS) \cite{Azure}. While \textit{Hadar} considers the task-level heterogeneity of DNN training jobs for scheduling decisions, Gavel only focuses on job-level heterogeneity, and Tiresias is heterogeneity-unaware among accelerators. For comparison, Tiresias is configured with two priority queues and its {\fontfamily{pcr}\selectfont PromoteKnob} disabled, whereas Gavel has its configuration similar to that under the previous experimental study \cite{gavel}. The comparative performance metrics of interest include GPU resource utilization (GRU) and the total time duration (TTD), with scalability also compared, as detailed next.

\renewcommand{\arraystretch}{1.25}
\begin{table}[t]
\centering
\hfill
\fontsize{10pt}{12pt}\selectfont 
\caption{Evaluation workloads: model, dataset, and relative size for each deep learning job}
\label{tab-models}
\vspace{20pt}
\resizebox{\columnwidth}{!}{
\begin{tabular}{|l|l|l|l|}
\hline
\textbf{Training Job }                                                           & \textbf{Model}                                                                                   & \textbf{Dataset}                            & \textbf{Size} \\ \hline
\begin{tabular}[c]{@{}l@{}}Image \\[-4pt]Classification\end{tabular} & ResNet-50 \cite{resnet50}                                                       & ImageNet \cite{imagenet}   & XL       \\ \hline
\begin{tabular}[c]{@{}l@{}}Image \\ [-4pt]Classification\end{tabular} & ResNet-18 \cite{resnet50}                                                       & CIFAR-10 \cite{cifar-10}   & S       \\ \hline
\begin{tabular}[c]{@{}l@{}}Language\\ [-4pt]Modeling\end{tabular}     & LSTM \cite{lstm}                                                            & Wikitext-2 \cite{wikitext} & L      \\ \hline
\begin{tabular}[c]{@{}l@{}} Image-to-Image \\[-4pt] Translation   \end{tabular}                                              & CycleGAN \cite{gan2} & Monet2photo \cite{gan2}     & M      \\ \hline
\begin{tabular}[c]{@{}l@{}}Language \\ [-4pt]Translation \end{tabular}                                           & Transformer \cite{transformer}                                                    & \begin{tabular}[c]{@{}l@{}}Multi30K \cite{multi30k}\\[-4pt](de-en) \end{tabular}    & L  \\      \hline

\end{tabular}
}
\end{table}
\renewcommand{\arraystretch}{1}

\begin{figure}
\centering
\resizebox{.4\textwidth}{0.2\textwidth}{%
\begin{tikzpicture}
\begin{axis}[
xlabel={GPU resource utilization (\%)},
ylabel={Fraction of time},
xmin=0, xmax=100,
ymin=0, ymax=1,
xtick={20,40,60,80,100},
ytick={.2,.4,.60,.80,1.00},
legend pos=north west,
ymajorgrids=true,
]

    \addplot[dash dot,thick,
        color=orange,
        ]
        coordinates {
           (0,0)(30,0.0074)(40,.0204)(50,.0223)(60,.0335)(70,.0335)(80,.05587)(90,.094)(95,.2)(97,1)};
  \addplot[dashed, thick,
      color=blue,
      ]
      coordinates {   (0,0)(30,.00616)(40,0.013)(50,.017)(60,.036)(70,.12)(80,.186)(90,.89)(93,1)};
\addplot[dotted, thick,
    color=green,
    ]
    coordinates {
    (0,0)(30,0.078)(40,0.028)(50,0.031)(60,0.075)(70,0.194)(80,0.402)(90,.913)};
    
    \addplot[dashed, thick,
    color=red,
    ]
    coordinates {
    (0,0)(33,0.0084)(38,.0169)(50,.0302)(73.333,.0326)(80,.0532)(93,.065)(96.6667,.214)(98.333,.2249)(96.6667,.23458)(98.333,.27206)(96.667,.37122)(95,.3954)(96.667,.4099)(95,.4147)(96.667,.5211)(93.333,.5320)(96.667,.6457)(90,.6578)(91.6667,.70737)(93.3333,.92987)(91.6667,.9807)(90,1)};
  \legend{\textit{Hadar},Gavel,Tiresias, YARN-CS}
\end{axis}
\end{tikzpicture}
}
\caption{Comparison of cluster-wide GPU resource utilization (GRU) among the four schedulers.}
\label{gpuutilization2}
\vspace{5pt}
\end{figure}
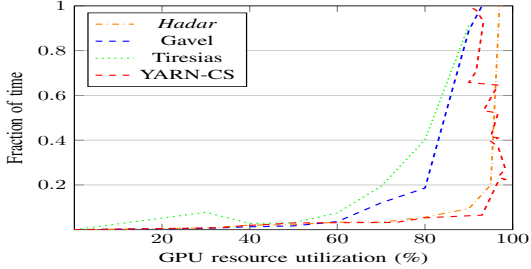

\begin{figure}[t]
\centering
\includegraphics[width=0.45\textwidth, height=0.20\textwidth]{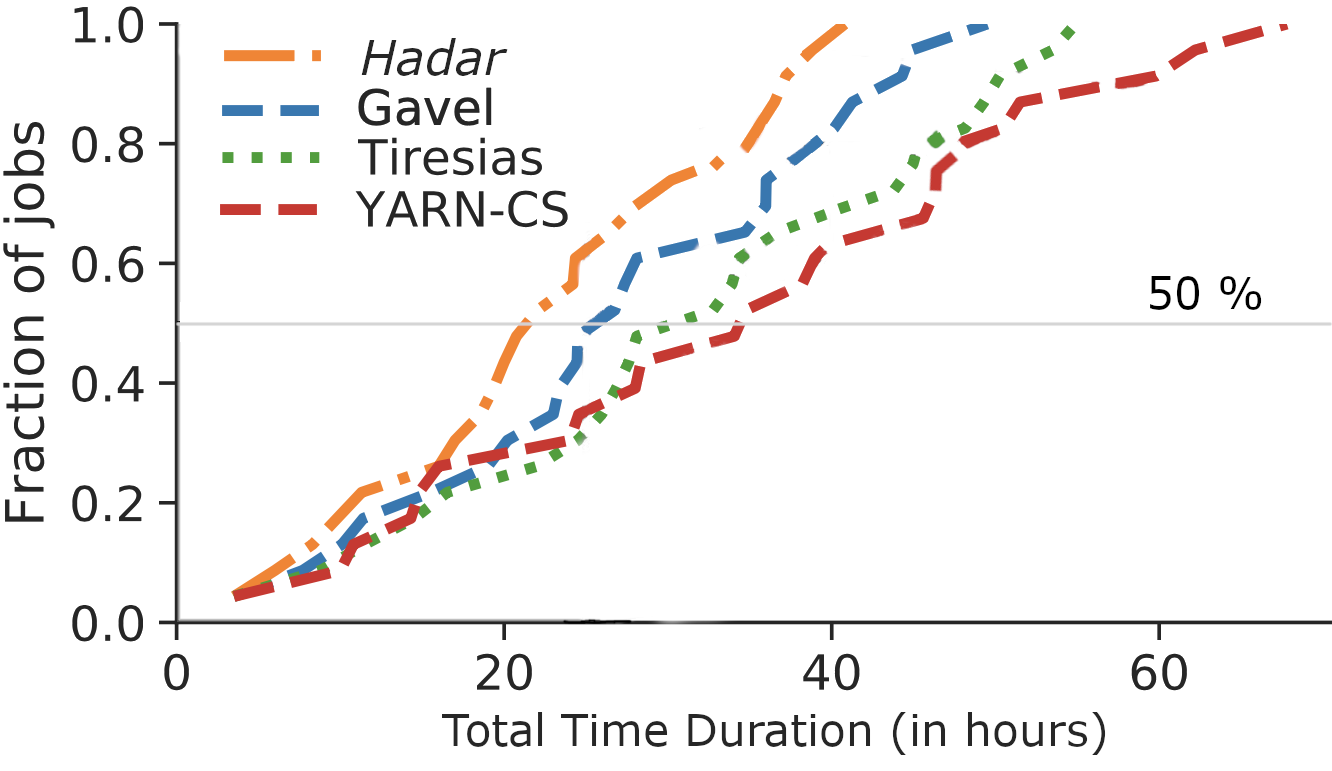}
\caption{Cumulative fractions of completed jobs over time, when scheduled by Gavel \cite{gavel}, Tiresias \cite{gu2019tiresias}, YARN-CS \cite{Azure}, and {\em Hadar}, respectively.}
\label{fig-jctvsfracjob}
\vspace{10pt}
\end{figure}

\textbf{GPU Resource Utilization.} 
GPU resource utilization (GRU) refers to the percentage of the total job run-time during which GPUs are utilized. The comparative GRU results of four schedulers are shown in Fig.~\ref{gpuutilization2}. The highest GRU is achieved by YARN-CS due to its non-preemptive nature. However, this comes at the cost of long total job completion duration, as to be seen in Fig.~\ref{fig-jctvsfracjob}.
Gavel, on the other hand, leaves heterogeneous GPUs unused even if the total number of them meets the requirement of a queued job. This results in lower GRU compared to that of YARN-CS. Tiresias also suffers from the same limitation as Gavel. %as it is not aware of the heterogeneity among accelerators.
In contrast, \textit{Hadar} takes advantage of fine-grained scheduling and resource heterogeneity awareness to elevate its GRU. More specifically, \textit{Hadar} can allocate tasks to GPUs that are most suited for them, possibly of different types when necessary, based on task characteristics and cluster resource availability. As a result, the number of GPUs left unused is minimized, leading to better GRU compared to those of Gavel and Tiresias. Moreover, \textit{Hadar} exhibits similar GRU compared to YARN-CS, indicating that it can utilize GPUs effectively to lower job completion times.

\textbf{Total Time Duration.} 
Cumulative fractions of completed jobs over time, when scheduled by Gavel \cite{gavel}, Tiresias \cite{gu2019tiresias}, YARN-CS \cite{Azure}, and \textit{Hadar}, are demonstrated in Fig.~\ref{fig-jctvsfracjob}, where \textit{Hadar} completes training all jobs in 40 hours, known as the total time duration (TTD). From the figure, \textit{Hadar} is observed to outperform its counterparts, whose TTDs equal ~68 hours by 1.67$\times$ compared to YARN-CS, and by 1.35$\times$ and 1.21$\times$ against Tiresias and Gavel, respectively. Additionally, the median time duration to complete 50\% jobs (marked by the horizontal gray line in Fig.~\ref{fig-jctvsfracjob}) under \textit{Hadar} is 1.20$\times$ (or 1.40$\times$) shorter than that under Gavel (or under Tiresias). Clearly, \textit{Hadar} outperforms all its counterparts due to better resource utilization.

\begin{figure}[t]
\centering
\includegraphics[width=0.40\textwidth, height=0.15\textwidth]{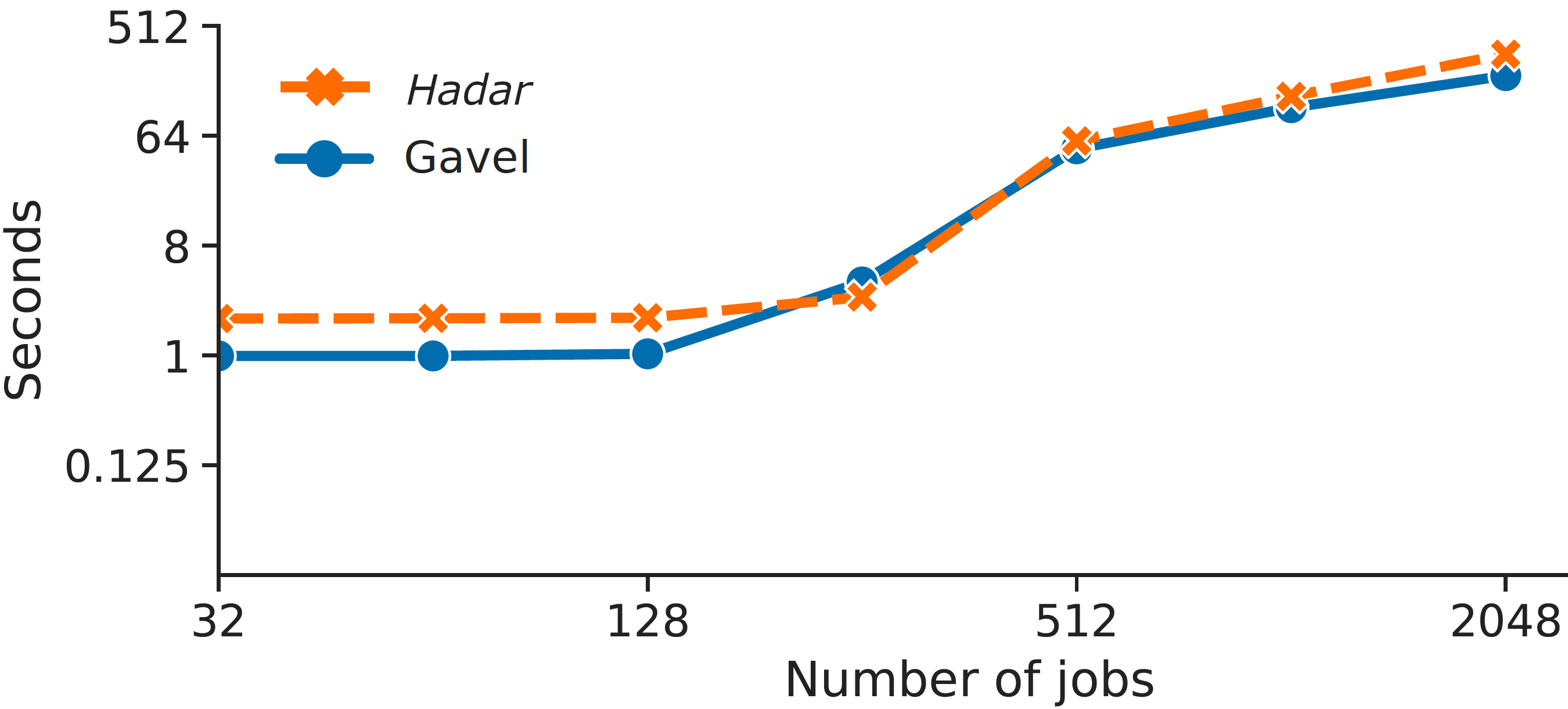}
\caption{Scalability comparison under {\em Hadar} and Gavel \cite{gavel} versus active jobs in a heterogeneous cluster, whose size grows as the number of jobs increases.}
\label{fig-load}
\vspace{10pt}
\end{figure}

\textbf{Scalability.} The time (in seconds) taken by {\em Hadar} and by Gavel to generate their decisions versus the number of jobs are depicted in Fig.~\ref{fig-load}. When the job count increases from 32 to 2048, {\em Hadar} and Gavel are observed to have similar scaling performance in terms of the scheduling time.
Even under heavy workloads (say, with some 2000 jobs), \textit{Hadar} can schedule, and allocate resources to, jobs in less than 7 minutes per scheduling round. %indicating that our algorithm scales well with increasing workload size. 
Our scheduler achieves resource allocation updating efficiently by dealing with solely newly incoming jobs in each scheduling round, if no preemption exists. Rather than recomputing the allocations of all jobs in every scheduling round, \textit{Hadar} just allocates resources to those newly incoming jobs progressively, without changing the allocated resources of running jobs present in the cluster. If the allocation of a running job is altered due to preemption, the affected job will get a new resource allocation, following a checkpoint/restart. We observe that only 30\% of scheduling rounds require changes to job resource allocations on average.

%Furthermore, we find that the scaling performance of our algorithm is quite similar to the scaling performance of Gavel.
\section{\textbf{Resource Utilization Enhancement}}
\label{sec-resource-utilization-enhancement}
While \textit{Hadar} is heterogeneous-aware to schedule DL training jobs run on a cluster across both spatial and temporal dimensions, each job is scheduled to run on at most one cluster node (i.e., machine or server) over training rounds. Hence, the cluster resource can often be under-utilized because a job cannot be executed concurrently on two or more nodes, even when some nodes are idle and available in any scheduling round. It is due to the fact that there is just one single copy of each job under training throughout all scheduling rounds, limiting \textit{Hadar} to allocate just one node at any job. As can be seen in Fig.~\ref{fig-dynamic-round}(a), \textit{Hadar} schedules three training DL jobs (J1, J2, and J3) to run on just three of the five nodes constituting the cluster testbed available in our lab, leaving two nodes idle for the first three rounds, from R1 to R3. Since J1 completes its training in R3, three nodes become idle during R4, when only two training jobs are in progress. The last four rounds involve just one node (Dell) to continue training J3 until completion. Our solution to overcoming this limitation is to fork a job to multiple copies, which may then be scheduled onto separate nodes available for training the job concurrently, yielding \textit{Hadar} Enhancement (or \textit{HadarE} for short). This way enhances resource utilization by avoiding cluster nodes to stay idle in any scheduling round, when there is a training job copy left to be completed. If a job is forked to \textit{f} copies, the job may be executed on up to \textit{f} nodes, if available. When the training times of jobs in a batch are known \textit{a priori}, it is desirable to fork these long-running jobs, each up to \textit{n} copies, for the best resource utilization under \textit{HadarE}. If each training job is forked to five copies when run on our 5-node testbed to keep all cluster nodes busy in all rounds but possibly the last one, as illustrated in Fig.~\ref{fig-dynamic-round}(b), highest resource utilization is achieved to yield the shortest TTD (total time duration).

Each training job in a workload batch under \textit{HadarE} is forked to \textit{n} copies for execution on an \textit{n}-node cluster, permitting multiple copies of a given job to run concurrently on up to \textit{n} nodes, if available. Two issues are involved in realizing \textit{HadarE} for the best performance: scheduling the copies of training jobs (plus initial throughput estimation) and aggregating and consolidating the training copies of each job.

\subsection{\textbf{Scheduling Copies of Training Jobs}}
\label{subsec-scheduling-forked-jobs}
\begin{figure*}[t]
    \centering
    \includegraphics[width=0.85\textwidth, height=0.18\textwidth]{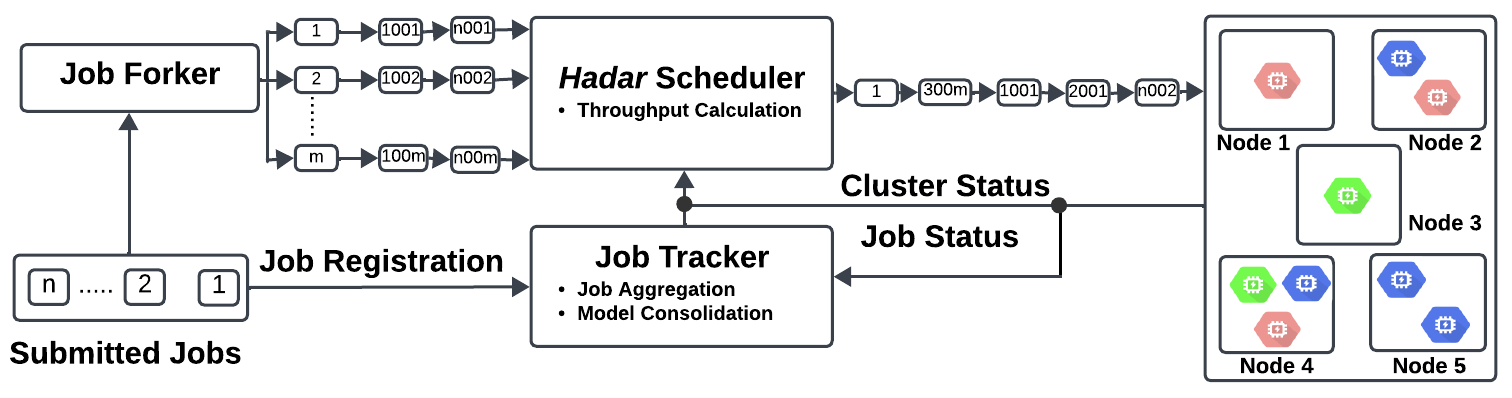}
    \caption{An overview of \textit{Hadar} Enhancement, {\em HadarE}.}
    \label{fig:block-diagram}
\end{figure*}

Each training job starts with forking it into a proper number of copies, say  \textit{n}, for an \textit{n}-node DN cluster. A Job Tracker is designed to track the progress of forked copies of all jobs, responsible for training copy aggregation and model parameter consolidation during the course of training, as shown in Fig.~\ref{fig:block-diagram}. All forked copies of a job are registered with Job Tracker, utilizing their unique job-IDs. Each job-ID is produced by
$\text{job\_ID} = \text{max\_job\_count} \times i + \text{parent\_job\_id}$, where \text{max\_job\_count} is the maximum number of jobs expected to co-exist in the \textit{n}-node cluster and \textit{i} ranges from 1 to the number of forked copies for the job, typically being \textit{n} to maximize resource utilization. With their unique job\_IDs obtained from Job Tracker, all copies of the job are sent to \textit{Hadar} for scheduling to run on cluster nodes (see Fig.~\ref{fig:block-diagram}).

\begin{figure}[t]
    \centering
    \begin{minipage}[b]{0.26\textwidth}
        \centering
        \includegraphics[width=0.9\textwidth]{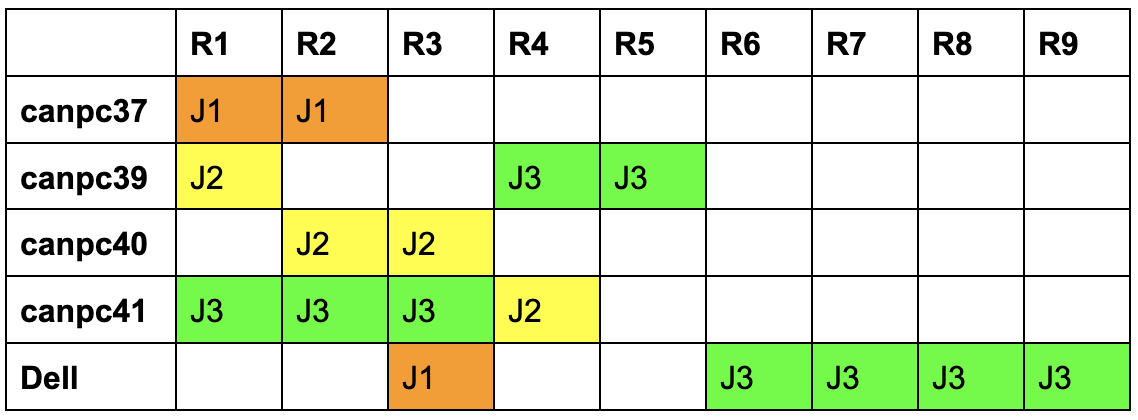}
        \caption*{(a) Under {\em Hadar}}
        \vspace{15pt}
        \label{fig:hadar-illustration}
    \end{minipage}
    \begin{minipage}[b]{0.21\textwidth}
        \centering
        \includegraphics[width=0.9\textwidth]{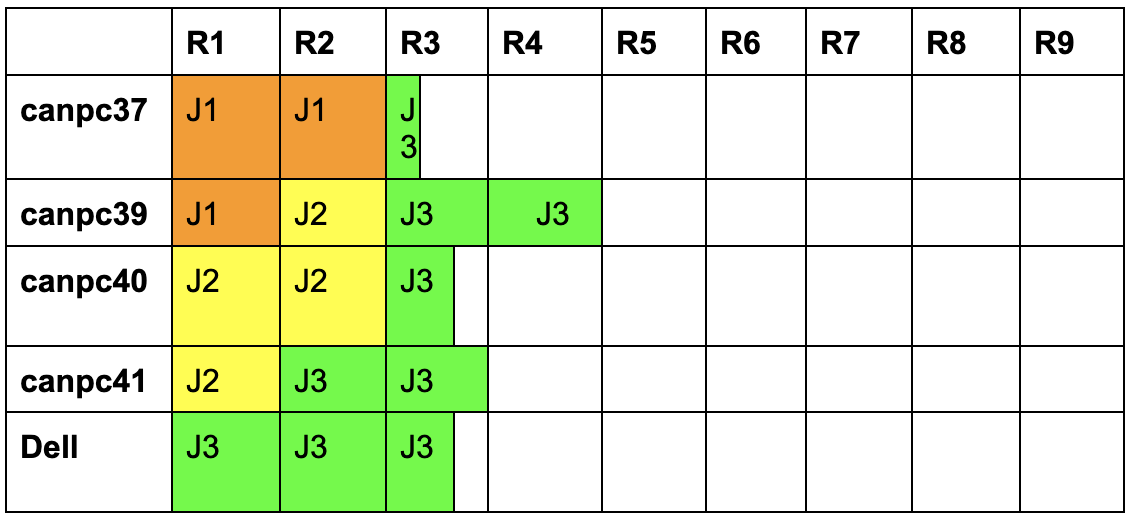}
        \caption*{(b) Under {\em HadarE}}
        \vspace{15pt}
        \label{fig:hadarE-illustration}
    \end{minipage}
\caption{Comparative illustration of scheduling rounds under (a) {\em Hadar} and (b) {\em HadarE}.}
\label{fig-dynamic-round}
\vspace{12pt}
\end{figure}

As demonstrated in Fig.~\ref{fig-dynamic-round}(b), \textit{HadarE} schedules the three jobs, with each of them forked to five copies (by Job Forker shown in Fig.~\ref{fig:block-diagram}) and registered with Job Tracker, for a total of 15 jobs to maximize resource utilization so that no node is left idle throughout the first two rounds. In the 3\textsuperscript{rd} round, all nodes are assigned to train J3 (the longest running job) for various dispatched epoch counts. The last round for one single node to finish up the remaining epochs of J3. In this figure, four copies of J3 complete their dispatched numbers of epochs at various time points before the scheduled R3 time slot expires, and the last round involves the node that completes its assigned epochs first in J3 training and is thus best suitable for handling J3's remaining epochs. In general, the last round may involve $\beta$ nodes (for 1 $\leq$ $\beta$ $\leq$ \textit{n}) to concurrently finish up the remaining epochs, with participating nodes being those $\beta$ fastest nodes that complete their assigned epochs in the immediately earlier round.  The job copy(s) run on the very last round for a batch of training jobs is (are) likely to end ahead of its scheduled time slot, exhibiting early finish of those training jobs. Clearly, \textit{HadarE} enjoys higher performance than \textit{Hadar}, boosting resource utilization to shorten the duration of training a batch of DL models and thus to lower the mean job completion time (JCT).
     
Relying on \textit{Hadar}, \textit{HadarE} schedules jobs in the round-based manner with a fix time slot per round.  During the time slot, each scheduled node trains its assigned job for the specified number of epochs.  The node may complete the specified number of epochs before the time slot expires; in this case, the node waits to get its next allocated job at the beginning of the next round. On the other hand, the node may fail to complete the specified number of epochs when the time slot ends; in this case, the node informs Job Tracker of the number of epochs it has completed, since this information is needed in scheduling the next round. Hence, communications take place between nodes and the tracker at each round after they complete their assigned epochs or the time slot expires, for the tracker to (1) calculate the total number of epochs completed by all allocated nodes for every job, (2) aggregate all copies of every job, and (3) consolidate the model parameters of every job obtained by its concurrently executed copies, where (2) and (3) are elaborated in subsection ~\ref{subsec-aggregation}.  As necessity, \textit{HadarE} augments the \textit{Hadar} Scheduler with initial throughput estimation (see Fig.~\ref{fig:block-diagram}), outlined below.

\textbf{Initial Throughput Estimation.} Following the primal-dual approach outlined in Section ~\ref{problem-primal-dual} for performance improvement, \textit{Hadar} depends on the measured speedups (i.e., throughputs) of each model when trained on all cluster nodes (which are characterized mainly by their equipped GPUs and associated on-board Performance-Memory Index (PMI), PCIe communication capacities, etc.).  Such throughput information can be either obtained by model execution profiling or given as the scheduling input. However, model execution profiling, even done on the fly, tends to take many rounds (so that every model has the chance to be assigned for execution on each node) before sound overall throughput information can be obtained. On the other hand, throughput information given as the input cannot be inclusive, so that any cluster containing nodes with different GPU and/or PCIe components absent in given throughput information would involve profiling in early rounds after scheduling starts. In both situations, unsatisfactory allocations happen from the beginning until the throughput of every model executed on each node is all available. To address this shortcoming, we have derived an expression for initial throughput estimation, after numerous experiments on the heterogeneous GPU testbed in our lab, as follows:
\begin{equation}
\text{Throughput} = \frac{PMI \times \text{batch\_size} \times \text{pcie\_scaling}}{\text{model\_weight} \times \text{dataset\_size}},
\end{equation} 
\noindent {\normalsize whose rationale is highlighted next.}

Since DL model training is often computation- and memory-intensive, best performed on GPUs (with tensor cores), PMI (Performance-Memory Index) in the expression denotes the ratio of GPU's parallel processing ability aided by tensor cores, in terms of teraflops (tera floating point operations) per second and the GPU's VRAM capacity in a square root fashion.  In the course of training on GPUs, the host machines of involved GPUs require frequent and heavy communications from host's DRAM and GPU's on-board VRAM through their Peripheral Component Interconnect Express (PCIe) channels, making PCIe capacities critical to GPU throughputs.  The expression thus includes the term of pcie\_scaling, which signifies different PCIe versions integrated on the machine's motherboard.  Additionally, the mini-batch size during training on a GPU affects its throughput, with a larger size calling for less communications to net a larger throughput.  On the other hand, a more complicated model or a larger dataset used for training on a GPU lowers its throughput, with model complexity approximated by a weight scale, from small, modest, high, to extra high; so is the dataset size to scale from S, M, L, to XL (see Table \ref{tab-models}).

Note that the aforementioned expression provides a reasonable estimate for \textit{HadarE} to yield good scheduling decisions from the beginning.  The quality of throughput information is improved progressively in the course of training, since every scheduled round let each involved node notify the Job Tracker of its actual throughput under its allocated job.  As a result, the accurate throughputs of every model on all nodes are progressively available.

\subsection{\textbf{Aggregating and Consolidating Training Copies}}
\label{subsec-aggregation}
Unlike \textit{Hadar} and earlier schedulers (Gavel, Tirsias, YARN-CS, etc.) where every training job is run on one node at a time throughout all scheduling rounds, \textit{HadarE} lets each training job be executed on multiple (available) nodes concurrently for resource utilization enhancement and thus training time reduction. It naturally needs to aggregate and consolidate results of copies of every job (i.e., DL model) trained on different nodes after each scheduled round to ensure the quality of models after training finishes, so that they are similar to, or even better than, what would have been obtained when trained without job forking. Both result aggregation and result consolidation are conducted by Job Tracker at the end of each scheduling round. Result aggregation simply sums completed training steps up, where one training step for a model means to train the model via $\phi$ mini-batches of training data, with $\phi$ = (training data size)/(mini-batch size). Note that in practice, model training progress is tracked at the step level, instead of the epoch level, especially when the scheduling time slot is short (i.e., a few minutes). When the total number of completed training steps of a job reaches the specified threshold, equal to \(\phi\)$\times$ (the training epoch count), all copies of the job are then discarded.

Upon completing its assigned training steps or ending the time slot, a node notifies Job Tracker of the number of steps it completed and the trained model parameters of its scheduled job. The total number of steps for the job completed by all nodes involved in training copies of the job is added (i.e., aggregated) by Job Tracker, with the job's model parameters consolidated by weight-averaging those of training copies, before passing them to the scheduler (\textit{Hadar}, see Fig.~\ref{fig:block-diagram}), which then makes allocations for the next round. Given the number of training steps of a job left at the start of a scheduling round, \textit{HadarE} for an \textit{n}-node cluster divides that number into \textit{n} portions according to their respective throughput values (that reflect nodes’ training capabilities), for assigning to those \textit{n} forked copies of the job. Every copy of the job, if scheduled to run on a node in the next round, will continue job training with its consolidated model parameters for its assigned number of steps.

\subsection{\textbf{Theoretical Analysis}}
\begin{theorem}[Maximal Resource Utilization]
\textit{HadarE} achieves the maximal cluster resource utilization of an \textit{n}-node cluster for training a batch of jobs forking its every job to \textit{n} copies for training.
\end{theorem}
\vspace{-10pt}
\noindent \textit{Proof.} Suppose there are $n$ nodes in a cluster running multiple jobs with the time slot per round of $T_{S}$ which is shorter than the shortest job's training time. The metric of interest, cluster resource utilization $(\text{CRU})^*$\footnotemark[0], for a single round, is the ratio of the total busy time spans during $T_{S}$ for all \textit{n} nodes to ($T_{S} \times$ \textit{n}), where the former is given by $\sum_{i=1}^{n} T_i^B$, with $T_i^B$ denoting the busy time span for node \textit{i}, $1 \le \textit{i} \le \textit{n}$, $\text{CRU} = \frac{1}{T_{S} \times n}\sum_{i=1}^{n} T_i^B.$ 

Our proof first considers one job which takes $R_1$ rounds, with the following four cases to be examined:
\footnotetext[0]{$^*$For clarity, $\text{CRU}_k^x$ refers to the CRU of $k$ total jobs (denoted by subscript), with each job forked to $x$ copies (denoted by superscript).} 

\textit{Case I: No job forking.} For only one job without job forking, \textit{HadarE} will reduce to \textit{Hadar} with only one of the nodes occupied in each round by the job. $\text{CRU}^1_1$ in this case is given by $\text{CRU}^1_1 = \frac{1}{R_1}\sum_{r=1}^{R_1} {\text{CRU}(r)},$
where $r$, $1 \le r \le R_1$, refers to the round number.

\textit{Case II: Job forking to x copies ($1 < x < n$).} With the job forked into \textit{x} copies, for $1 < x < n$, cluster resource utilization (denoted by $\text{CRU}^x_1$) then rises in a single round, as \textit{x} nodes (out of \textit{n} total nodes) will be running \textit{x} forked jobs. As \textit{x} copies of the job are trained simultaneously (before their obtained model parameters are consolidated, in preparation for the next training round), the total number of rounds to complete the job is reduced accordingly (to equal the ceiling function of the ratio $\frac{R_1}{x}$). The cluster resource utilization is therefore given by {\small $\text{CRU}^x_1 = \frac{1}{\lceil{R_1/x}\rceil} \sum_{r=1}^{\lceil{R_1/x}\rceil} {\text{CRU}(r)}$}.

\textit{Case III: Job forking to \textit{n} copies.} This case forks the job to \textit{n} copies, giving rise to $\text{CRU}^n_1 = \frac{1}{\lceil{R_1/n}\rceil}\sum_{r=1}^{\lceil{R_1/n}\rceil} {\text{CRU}(r)}$.

\textit{Case IV: Job forking to (n + j) copies.} Since only \textit{n} copies will be assigned to the nodes at a time, the additional forked job will not reduce the number of rounds needed to complete model training, and thus cluster resource utilization ($\text{CRU}^{n+j}_1$) is the same as $\text{CRU}^n_1$. By observing equations above, we have 
\begin{align}\label{equ-cru-conclusion}
    {\small\text{CRU}^1_1 < \text{CRU}^x_1 < \text{CRU}^n_1= \text{CRU}^{n+j}_1,}
\end{align} \noindent {\normalsize where \textit{j} refers}
to additional copies of forked jobs beyond \textit{n} copies.
Eq. (\ref{equ-cru-conclusion}) signifies that cluster resource utilization under a single job is maximum when the job is forked into \textit{n} copies, serving as the initial step of our proof.

Let's assume that Eq. (\ref{equ-cru-conclusion}) is true for $k\;(>1)$ jobs with $R_k$ total rounds and the CRU of $\text{CRU}_k$. We have
{\small
\begin{align}\label{equ-cru-conclusion-k}
    \text{CRU}_k^1 < \text{CRU}_k^x < \text{CRU}_k^n = \text{CRU}_k^{n+j}.
\end{align} \noindent {\normalsize Next, }
}
the derivation of $CRU_{k+1}$ under $k+1$ jobs is provided by considering the additional job, as follows. For simplicity, let the additional job be the last one. The job can then be viewed as what we examined above for the single job situation, with $R_1$ rounds to complete without forking. If the last job for scheduling is not the additional one, the same conjectural process holds.

For $k+1$ jobs forked into $x$ copies, $\text{CRU}^n_{k+1}$ and $\text{CRU}^x_{k+1}$ are similar when number of forked jobs is greater than or equal to number of nodes, $n \leq [(k+1) * x] $, but $\text{CRU}^x_{k+1}$ decreases significantly than $\text{CRU}^n_{k+1}$ for each additional round when $[(k+1) * x] < n$. Evidently, for the latter case, the total number of rounds to complete the remaining jobs when forked into $x$ copies is higher than when forked into $n$ copies, so the $\text{CRU}^x_{k+1}$ over the entire process reduces along with each additional round of the process. For example, assume only one job is remaining when $(k * x) \le n$, and it takes $R_n^{re}$ rounds to complete the remaining job when forked into $n$ copies, then the remaining job will complete in $\lceil \frac{{R_n^{re}} * n}{x} \rceil $ rounds when forked into $x$ copies. The CRU for the remaining rounds (without considering the overheads) is $\frac{x * 100}{n}$\% (or 100\%) when forked into $x$ (or $n$) copies. Thus, $\text{CRU}^n_{k+1}$ is higher than $\text{CRU}^x_{k+1}$. Similarly, for jobs without forking, when number of jobs is less than number of nodes ($(k+1) < n$), the CRU in each round (without considering overhead) is $\frac{1 * 100}{n}$\%. With the aforementioned fact about the CRUs of $k+1$ jobs under different cases (without forking, every job forked into $x$ copies, every job forked into $n$ copies, and every job forked into $x+j$ copies) and the facts of Eqs. (\ref{equ-cru-conclusion}) and (\ref{equ-cru-conclusion-k}), we infer that CRU is smaller without job forking than with every job forked into \textit{x} copies and that the maximum CRU is obtained when each job is forked into \textit{n} copies, arriving at
\begin{align}\label{equ-cru-conclusion-k1}
    {\small \text{CRU}_{k+1}^1 < \text{CRU}_{k+1}^x < \text{CRU}_{k+1}^n = \text{CRU}_{k+1}^{n+j}.}
\end{align} \noindent {\normalsize Since Eq. (\ref{equ-cru-conclusion}),}
the base case, is true, and Eq. (\ref{equ-cru-conclusion-k}) is assumed to be true for proving Eq. (\ref{equ-cru-conclusion-k1}), we conclude that the maximum CRU, is attained when each job is forked to \textit{n} copies, yielding
\begin{align}\label{equ-cru-conclusion-n}
    {\small \text{CRU}^1_{m} < \text{CRU}^x_{m} < \text{CRU}^n_{m} = \text{CRU}^{n+1}_{m},}
\end{align} \noindent {\normalsize where $m$ is any number of jobs in a batch. \hfill $\qedsymbol$}

\noindent\textbf{Corollary:}  Under \textit{HadarE} with every job forked to \textit{n} copies in an \textit{n}-nodes cluster, no idle node exists in any scheduling round but possibly the last one. 

For every job, there are \textit{n} forked jobs. Each of the \textit{n} copies can be assigned to one of the nodes. So, every node will get a forked job to complete in each round. Nodes will never be idle if all nodes are needed to complete remaining jobs. Hence, only the last round of the entire training course may a node be idle, and any earlier round will have every node assigned with one remaining job.

\section{\textbf{Evaluation on Physical Clusters}}
\label{sec:physical-cluster-eval}

% In this section, we conduct extensive experiments on physical GPU clusters to evaluate the performance of \textit{HadarE} in terms of the job completion time (JCT), GPU utilization, and total duration compared to state-of-the-art schedulers such as \textit{Hadar} and Gavel. {\em HadarE} is evaluated on both testbeds available in our lab and clusters in Amazon Web Services (AWS). The evaluation metrics being employed are Average Job Completion Time (JCT), total duration of training, and finally cluster utilization against the parent scheduler \textit{Hadar} and one of the state-of-the-art schedulers, Gavel. The round-time for all the schedulers is set to 360 seconds. \textit{Hadar} performs best at this round-time slot \cite{hadar}. 

Extensive experiments are conducted to evaluate \textit{HadarE} on physical GPU clusters either leased from the AWS (Amazon Web Services) Cloud (called the AWS cluster) or available at our research lab (called the testbed cluster), for comparison with \textit{Hadar} and Gavel.

\subsection{\textbf{Experimental Setup}}
\label{subsec-experimental-setup}
The AWS physical cluster comprises five nodes located in the same AWS region to keep the network latencies low, including one p3.2xlarge node equipped with a Tesla v100 GPU having 16 GB on-board VRAM, two p2.xlarge nodes each equipped with a Tesla K80 GPU having 12 GB on-board VRAM, and two g4dn.xlarge nodes each equipped with a Tesla T4 GPU having 16 GB on-board VRAM. Meanwhile, the heterogeneous testbed cluster available in our lab includes five nodes equipped respectively with Nvidia Titan RTX GPUs (24 GB VRAM), Tesla T4 GPU (16 GB VRAM), Nvidia T400 GPU (4 GB VRAM), GeForce RTX 3090 GPUs (24 GB VRAM), and Nvidia RTX A2000 GPU (6 GB VRAM). While some cluster nodes have two GPUs each, our evaluation always utilizes one GPU for every node.

\subsection{\textbf{Workloads and Datasets}}
\label{subsec-workloads-datasets}
Experiments are undertaken under five distinct DL models, with four of them present in a Microsoft trace \cite{microsoft-trace} and the fifth for predicting weather parameters, called the Encoder-Decoder Transformer model ~\cite{Mima} ~\cite{mima-codes}.  Those DL models  are summarized in Table ~\ref{tab-models-hadarE}, and they cover different applications with their dataset sizes ranging from S (small) to XL (extra large), same as trace-driven evaluation. The experiments are performed under batches of seven workload mixes, which involve various numbers of DL models, ranging from 1 to 12. Specifically, workload mix-1 (M-1) involves just one MiMa weather prediction model, whereas workload mix-3 (M-3) contains one Language Translation model and two MiMa models, denoted by $<$LT, 2$\times$MM$>$. Other five workload mixes are: M-4 = $<$IC, LM, LT, MM$>$, M-5 = $<$IC, LM, LT, RS, MM$>$, M-8 = $<$IC, LM, LT, RS, 4$\times$MM$>$, M-10 = $<$IC, LM, LT, RS, 6$\times$MM$>$, and M-12 = $<$IC, LM, LT, RS, 8$\times$MM$>$.

\subsection{\textbf{Evaluation Results}}
\label{subsec-results}
We conducted experiments to compare \textit{Hadar} and \textit{HadarE} against the best previous DL scheduler, Gavel \cite{gavel}, in terms of performance metrics of interest, including cluster resource utilization (CRU), the total time duration (TTD) taken to complete a batch of jobs, and the average job completion time (JCT) of all jobs. 

%\subsection{Simulation Fidelity}We mimic the workload used in our testbed experiments in the simulator to verify the fidelity of our simulator. Weobserve the simulation results to be similar to that of our testbed results ??$\times$ average JCT improvement w.r.t. Gavel, ?? $\times$  w.r.t. Tiresias, and ?? $\times$ w.r.t. YARN-CS. As simulation can not exactly capture overheads of preemption, cluster dynamics and the impact of placement the results are different to a small extent.
\renewcommand{\arraystretch}{1.25}
\begin{table}[t]
\hfill
\centering
\fontsize{10pt}{12pt}\selectfont 
\caption{Details of five DL models employed to construct workload batches run on physical clusters for evaluation}
\label{tab-models-hadarE}
\vspace{20pt}
\resizebox{\columnwidth}{!}{

\begin{tabular}{|l|l|l|l|}
\hline
\textbf{Training Job}                                    & \textbf{Model}                                                                                   & \textbf{Dataset}                            & \textbf{Size} \\ \hline

\begin{tabular}[c]{@{}l@{}}Image \\ [-4pt]Classification (IC) \end{tabular} & ResNet-18 \cite{resnet50}                                                       & CIFAR-10 \cite{cifar-10}   & S       \\ \hline

\begin{tabular}[c]{@{}l@{}}Language\\[-4pt] Modeling (LM) \end{tabular}     & LSTM \cite{lstm}      
& Wikitext-2 \cite{wikitext} & L      \\ \hline

\begin{tabular}[c]{@{}l@{}}Language \\[-4pt]Translation (LT) \end{tabular}                                           & Transformer \cite{transformer}                                                    & \begin{tabular}[c]{@{}l@{}}Multi30K \cite{multi30k}\\(de-en) \end{tabular}    & L  \\      \hline
\begin{tabular}[c]{@{}l@{}}Recommendation \\ [-4pt]System (RS)\end{tabular} & Recorder \cite{recoder}                                        & ML-20M \cite{ml-20m}   & XL       \\ \hline

\begin{tabular}[c]{@{}l@{}} MiMa Weather \\[-4pt] Predictions (MM)   \end{tabular}                                              & \begin{tabular}[c]{@{}l@{}}Encoder-Decoder  \\ [-4pt]Transformer \cite{Mima} \cite{mima-codes} \cite{zhang2024regional} \end{tabular} & \begin{tabular}{@{}l@{}}Mesonet \cite{kymesonet} and   \\ [-4pt]WRF-HRRR \cite{hrrr} \end{tabular}    & M      \\ \hline
\end{tabular}
}
\end{table}
\renewcommand{\arraystretch}{1}

\begin{figure*}[h]
    \centering
    \begin{minipage}{0.49\textwidth}
        \centering
        \includegraphics[width=0.90\textwidth, height=0.42\textwidth]{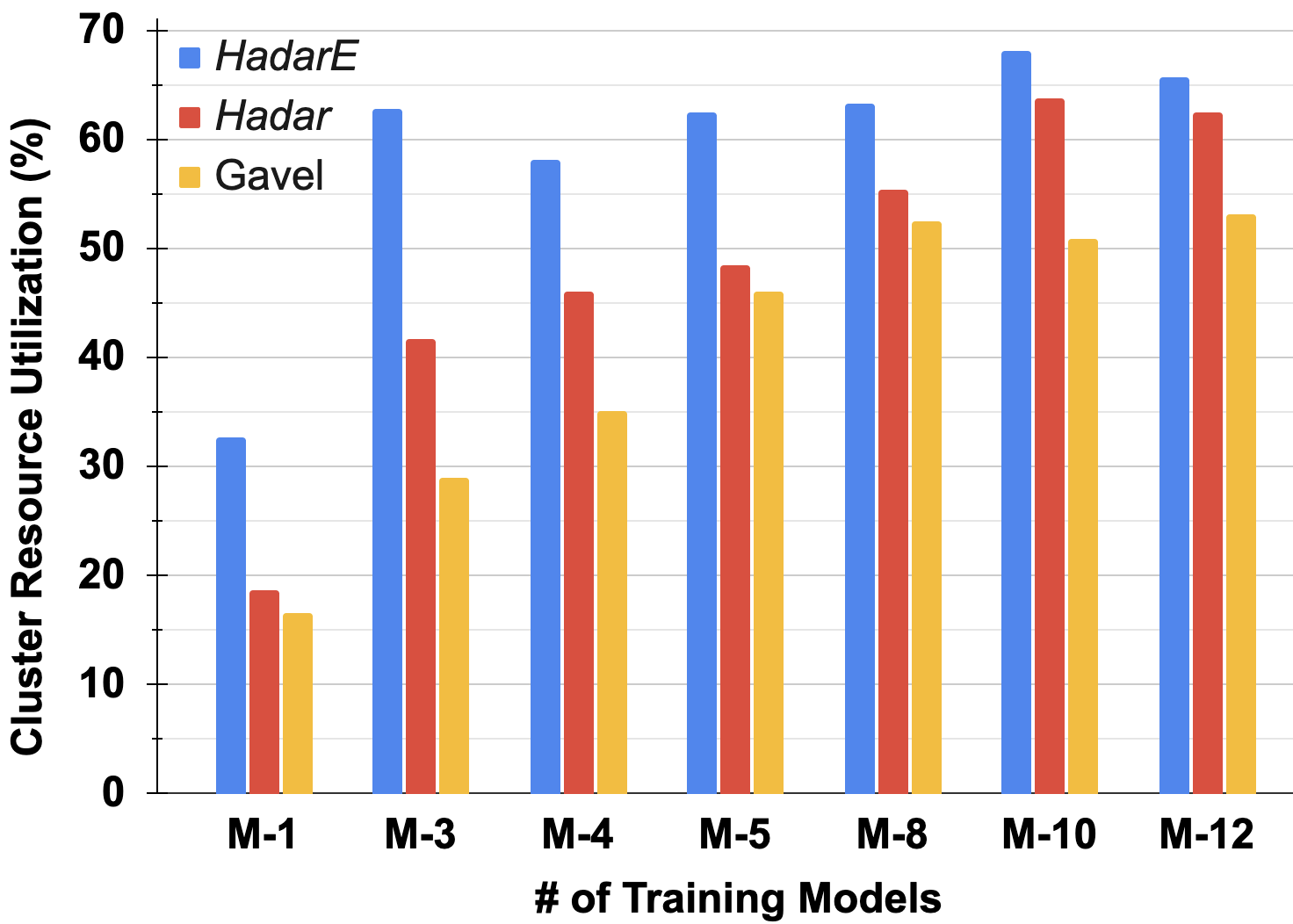}
        \caption*{(a) Cluster Resource Utilization (CRU) for AWS Cluster}
        \vspace{15pt}
        \label{fig:cu-aws}
    \end{minipage}
    \hfill
    \begin{minipage}{0.49\textwidth}
        \centering
        \includegraphics[width=0.90\textwidth, height=0.42\textwidth]{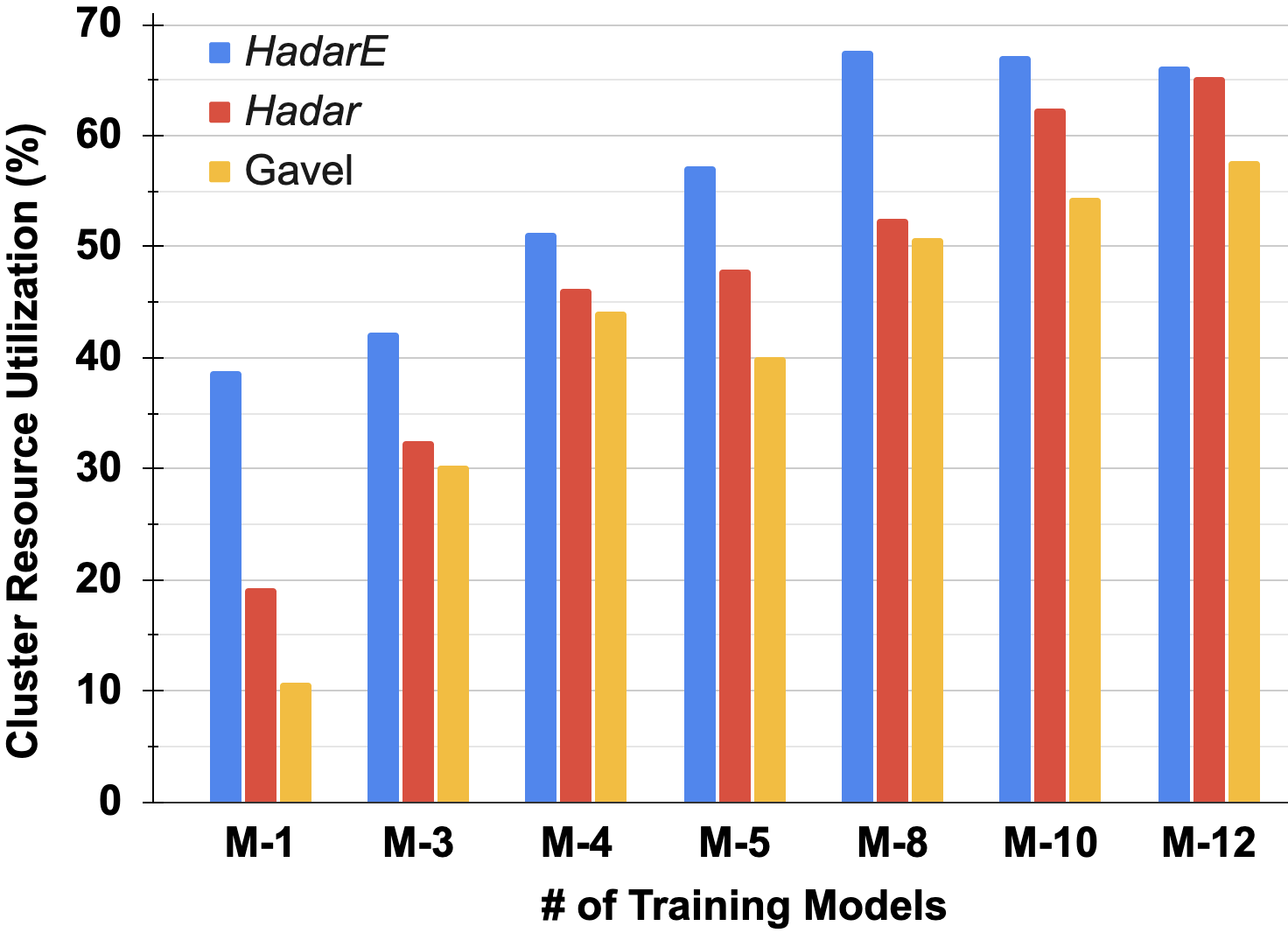}
        \caption*{(b) Cluster Resource Utilization (CRU) for Testbed Cluster}
        \vspace{15pt}
        \label{fig:cu-physical}
    \end{minipage}
\caption{Comparison of cluster resource utilization among Gavel \cite{gavel}, {\em Hadar} \cite{hadar}, and {\em HadarE} for AWS and testbed clusters.}
\label{fig:cluster-utilization}
\vspace{0.45cm}
\end{figure*}
\vspace{5pt}
\begin{figure*}[h]
    \centering
    \begin{minipage}[b]{0.49\textwidth}
        \centering
        \includegraphics[width=0.90\textwidth, height=0.42\textwidth]{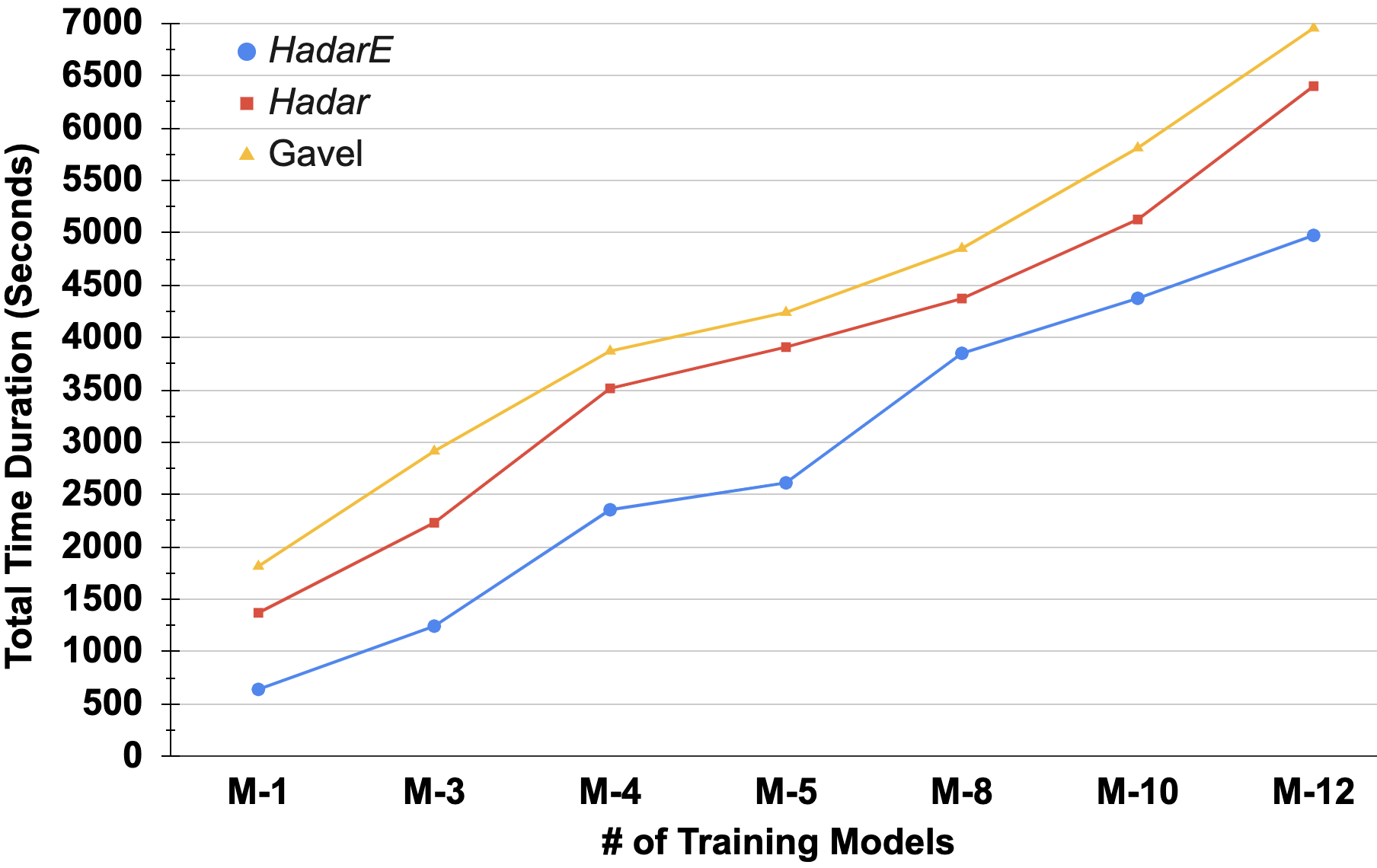}
        \caption*{(a) Total Time Duration (TTD) for AWS cluster}
        \vspace{15pt}
        \label{fig:td-aws}
    \end{minipage}
    \hfill
    \begin{minipage}[b]{0.49\textwidth}
        \centering
        \includegraphics[width=0.90\textwidth, height=0.42\textwidth]{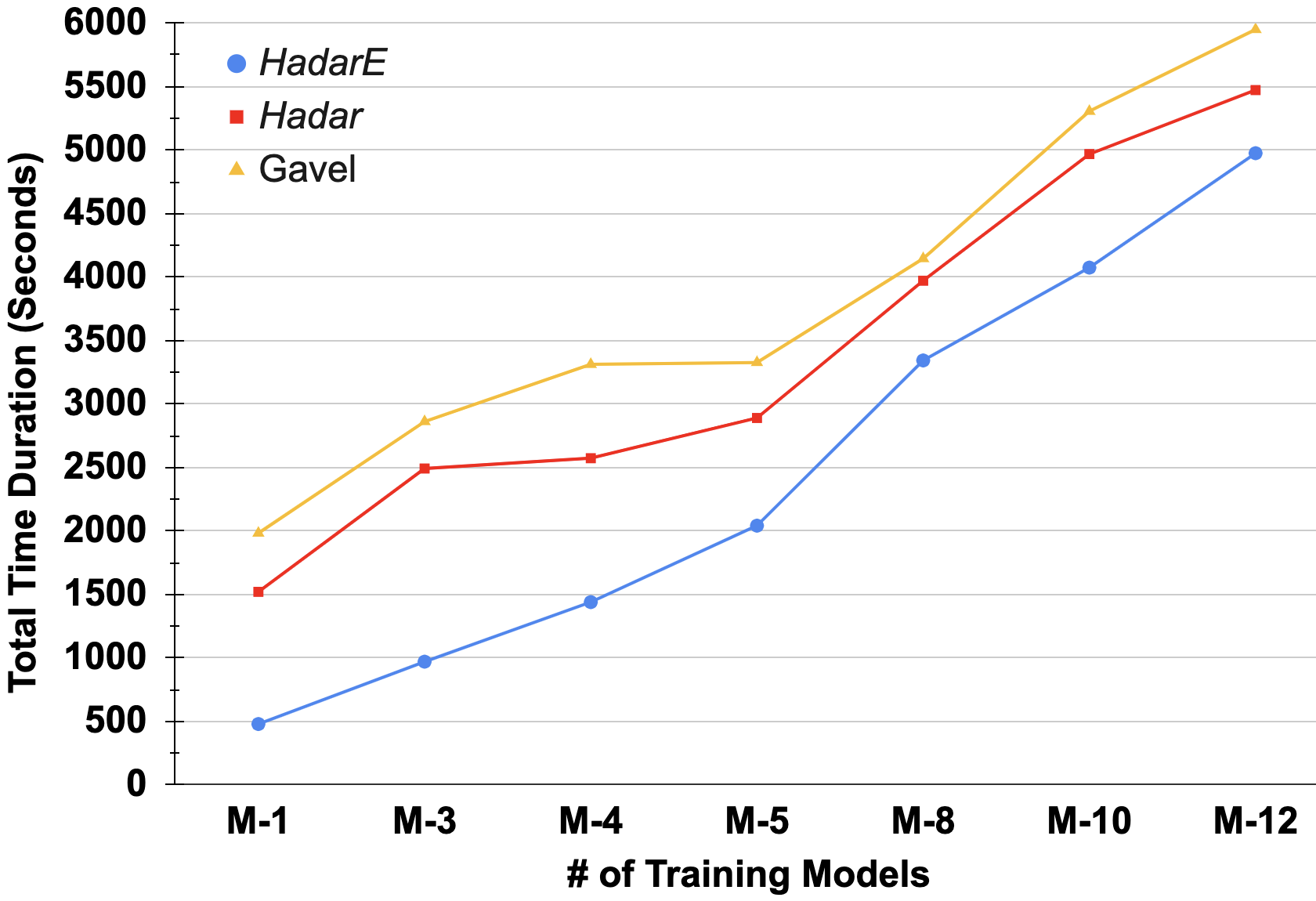}
        \caption*{(b) Total Time Duration (TTD) for testbed cluster}
        \vspace{15pt}
        \label{fig:td-physical}
    \end{minipage}
\caption{Comparison of TTD among Gavel \cite{gavel}, \textit{Hadar} \cite{hadar}, and \textit{HadarE} for AWS and testbed clusters.}
\label{fig:total-duration}
\vspace{5pt}
\end{figure*}

\textbf{Cluster Resource Utilization (CRU).} CRU refers to the ratio of the total busy times of all cluster nodes over the allocated time slots of all nodes. Fig.~\ref{fig:cluster-utilization} compares the resource utilization result of three schedulers, Gavel, \textit{Hadar}, and \textit{HadarE}. As evident by the Fig.~\ref{fig:cluster-utilization} \textit{Hadar} exhibits greater cluster resource utilization than Gavel, enjoying cluster utilization gain of 1.20$\times$ (or 1.21$\times$) in comparison with Gavel on the AWS (or testbed) cluster, as can be observed in \ref{fig:cluster-utilization}(a) (or \ref{fig:cluster-utilization}(b)). \textit{HadarE} achieves a 1.56$\times$ (or 1.62$\times$) performance gain on average against Gavel on AWS (or testbed). Apparently, the CRU gain of \textit{HadarE} is dictated by the difference between the number of training jobs left and the cluster node count. For example, if just one job is left to be trained on a 5-node cluster, four nodes are idle under all schedulers but \textit{HadarE}, which lets five copies of the remaining job run concurrently on all nodes to get the highest CRU utilization possible.

\label{total-duration}
\textbf{Total Time Duration (TTD).} The TTD results of seven different mixes of workloads under AWS and testbed clusters are illustrated in Fig.~\ref{fig:total-duration}. From the figure, it is clear that \textit{Hadar} outperforms its counterpart, Gavel, to have smaller TTDs to complete all training job(s) under every workload mix on both physical clusters. For example, Gavel takes some 4200 seconds (or 3300 seconds) to finish the M-5 workload mix versus 3900 seconds (or 2800 seconds) required by \textit{Hadar} on the AWS (or our testbed)  luster, to enjoy the training speedup of 1.18$\times$ (or 1.15$\times$), as shown in Fig.~\ref{fig:total-duration}(a) (or Fig.~\ref{fig:total-duration}(b)). For all the seven workload mixes, \textit{Hadar} on average exhibits the speedup of 1.17$\times$ (or 1.16$\times$) on the AWS (or testbed) cluster, in comparison to Gavel. Note that the performance gaps between \textit{Hadar} and Gavel change insignificantly across al workload mixes, because if a given mix sees the cluster node(s) idle under Gavel, the mix is expected to idle the cluster node(s) under \textit{Hadar} as well.

As expected, \textit{HadarE} achieves further performance gains against \textit{Hadar} (or Gavel), to yield the mean speedup of 1.79$\times$ (or 2.12$\times$) over all seven workload mixes. It is due mainly to forking every training job to multiple copies so that a job can be trained on multiple nodes simultaneously for shortening the TTD. This way lets \textit{HadarE} achieve more pronounced performance gains for workload mixes that cause Gavel and \textit{Hadar} to idle cluster nodes.

\begin{figure*}[h]
    \centering
    \begin{minipage}[b]{0.49\textwidth}
        \centering
        \begin{tikzpicture}
            \begin{axis}[
                ylabel=\scriptsize \textbf{Average Job Completion Time (Seconds)},
                xlabel= \scriptsize \textbf{\# of Training Models},
                ybar, % Changed from ybar interval to ybar
                ymin=250, % Start y-axis from 0
                ymax=5500, % Set the maximum value of the y-axis
                % ytick={1000,2000,3000,4000, 5000,6000}, % Set the y-axis tick intervals
                ytick={500,1500,2500,3500, 4500,5500}, % Set the y-axis tick intervals
                yticklabel style={font=\bfseries},
                symbolic x coords={1, 3, 4, 5, 8, 10, 12}, % Define the specific x-coordinates
                xtick=data, % Ensure all defined x-coordinates are used as ticks,
                xticklabels={M-1, M-3, M-4, M-5, M-8, M-10, M-12},
                xticklabel style={font=\bfseries},
                 % making the scales bold
                legend style={at={(0,1)}, anchor=north west},
                bar width=6pt,
                width=0.84\textwidth,
                height=0.54\textwidth
                ]
                \addplot+[error bars/.cd, y dir=both, y explicit]
                coordinates {
                    (1,800.7) += (0,0) -= (0, 0)
                    (3,1004.965) += (0, 178.384) -= (0, 89.305)
                    (4,1120.623) += (0, 571.469) -= (0, 286.829)
                    (5,1032.514) += (0, 565.528) -= (0, 348.874)
                    (8,1148.918) += (0, 928.374) -= (0, 204.938)
                    (10,1127.34) += (0, 491.652) -= (0, 298.339)
                    (12,1143.013) += (0,329.887) -= (0, 376.368)
                };
        
                \addplot+[error bars/.cd, y dir=both, y explicit]
                coordinates {
                    (1,1269.036) += (0, 0) -= (0, 0)
                    (3,1801.681) += (0, 1078.285) -= (0, 698.276)
                    (4, 1848.652) += (0, 1162.961) -= (0, 753.067)
                    (5, 1727.541) += (0, 1190.503) -= (0, 517.524)
                    (8,2277.292) += (0, 1138.46) -= (0, 1141.698)
                    (10,2589.764) += (0, 992.178) -= (0, 766.039)
                    (12,2653.107) += (0, 1493.452) -= (0, 835.908)
                };
        
                \addplot+[error bars/.cd, y dir=both, y explicit]
                coordinates {
                    (1,1728.573) += (0, 0) -= (0, 0)
                    (3,2036.155) += (0, 828.462) -= (0, 367.992)
                    (4, 2024.074) += (0, 1366.05) -= (0, 821.438)
                    (5, 2128.382) += (0, 1031.225) -= (0, 845.106)
                    (8,2526.593) += (0, 917.761) -= (0, 1167.922)
                    (10,2849.106) += (0, 1203.408) -= (0, 1528.877)
                    (12,3209.305) += (0, 2153.853) -= (0, 1169.569)
                };
                \legend{{\em HadarE}, {\em Hadar}, Gavel}
            \end{axis}
        \end{tikzpicture}
        \caption*{ (a) Average Job Completion Time for AWS cluster}
        \vspace{15pt} % Adjust the space as needed
        \label{fig-jct-aws}
    \end{minipage}
    \hfill
    \begin{minipage}[b]{0.49\textwidth}
        \centering
        \begin{tikzpicture}
            \begin{axis}[
                ylabel=\scriptsize \textbf{Average Job Completion Time (Seconds)},
                xlabel= \scriptsize \textbf{\# of Training Models},
                label style={font=\scriptsize},
                ybar, % Changed from ybar interval to ybar
                ymin=250, % Start y-axis from 0
                ymax=6000, % Set the maximum value of the y-axis
                ytick={1000,2000,3000,4000, 5000,6000}, % Set the y-axis tick intervals
                symbolic x coords={1, 3, 4, 5, 8, 10, 12}, % Define the specific x-coordinates
                xtick=data, % Ensure all defined x-coordinates are used as ticks,
                xticklabels={M-1, M-3, M-4, M-5, M-8, M-10, M-12},
                xticklabel style={font=\bfseries},
                yticklabel style={font=\bfseries}, % making the scales bold
                legend style={at={(0,1)}, anchor=north west},
                bar width=6pt,
                width=0.84\textwidth,
                height=0.54\textwidth
            ]
                \addplot+[error bars/.cd, y dir=both, y explicit]
                coordinates {
                    (1,459.56) += (0, 0) -= (0, 0)
                    (3,702.22) += (0, 237.745) -= (0, 237.499)
                    (4,730.5) += (0, 542.764) -= (0, 97.935)
                    (5,755.92) += (0, 807.472) -= (0, 216.833)
                    (8,943.98) += (0, 271.147) -= (0, 149.482)
                    (10,1127.34) += (0, 214.99) -= (0, 341.178)
                    (12,1202.987) += (0, 437.253) -= (0, 393.123)
                };
        
                \addplot+[error bars/.cd, y dir=both, y explicit]
                coordinates {
                    (1,1308.527) += (0, 0) -= (0, 0)
                    (3,1739.97) += (0, 350.442) -= (0, 462.078)
                    (4, 1613.265) += (0, 414.364) -= (0, 694.375)
                    (5, 1521.99) += (0, 848.125) -= (0, 661.207)
                    (8, 2150.48) += (0, 448.218) -= (0, 1103)
                    (10, 2256.382) += (0, 1338.667) -= (0, 1071.104)
                    (12, 2512.483) += (0, 2215.493) -= (0, 1486.034)
                };
        
                \addplot+[error bars/.cd, y dir=both, y explicit]
                coordinates {
                    (1,1743.092) += (0, 0) -= (0, 0)
                    (3,1913.32) += (0, 722.56) -= (0, 1089.682)
                    (4, 2168.377) += (0, 907.067) -= (0, 1398.267)
                    (5, 2178.659) += (0, 826.019) -= (0, 1218.048)
                    (8, 2184.726) += (0, 1760.225) -= (0, 736.02)
                    (10, 2588.372) += (0, 2407.021) -= (0, 965.766)
                    (12, 2814.906) += (0, 2515.399) -= (0, 1125.493)
                };
                \legend{\em{HadarE}, \em{Hadar}, Gavel}
            \end{axis}
        \end{tikzpicture}
        \caption*{(b) Average Job Completion Time for testbed cluster} % Adjust the space as needed
        \label{fig-jct-physical}
        \vspace{15pt}
    \end{minipage}
    \caption{Comparison of JCT among Gavel \cite{gavel}, {\em Hadar} \cite{hadar}, and {\em HadarE} for AWS and testbed clusters.}
    \label{fig-jct}
\end{figure*}
\textbf{Average Job Completion Time (JCT).} As demonstrated in Fig.~\ref{fig-jct}, \textit{Hadar} consistently exhibits smaller JCT values than its previous counterpart, Gavel, for all seven workload mixes (with 1 to 12 training models) under both AWS and testbed clusters. Given the workload with 5 training models (M-5), for example, JCT is 1.23$\times$ (or 1.43$\times$) smaller for \textit{Hadar} than for Gavel under the AWS (or testbed) cluster.  The maximal JCT and the minimal JCT for a given workload mix (with more than one training model) and a scheduler are indicated by a range mark on its average JCT bar. For the workload mix with five training models, as an example, the maximal JCT and the minimal JCT on the AWS cluster under \textit{Hadar} equal 2918 seconds and 1210 seconds, respectively, whereas they under Gavel are 3159 seconds and 1283 seconds, according to Fig.~\ref{fig-jct}(a). Across all seven workload mixes, \textit{Hadar} enjoys 1.17$\times$ (or 1.23$\times$) shorter JCT on average, when compared with Gavel, under the AWS (or testbed) cluster, according to Fig.~\ref{fig-jct}(a) (or Fig.~\ref{fig-jct}(b)). As expected, \textit{HadarE} exhibits further JCT reduction in comparison to Gavel, yielding JCT reduction by 2.23$\times$ (or 2.76$\times$) under the AWS (or testbed) cluster.  It also enjoys substantial JCT reduction, in comparison with \textit{Hadar} for both AWS and testbed clusters, due to its effective boost in resource utilization. In fact, the reduction degrees from \textit{HadarE} to \textit{Hadar} are observed to be consistently far larger than those from \textit{Hadar} to Gavel, under both clusters.  Interestingly, the JCT range of every workload mix (with more than one training model) is more confined under \textit{HadarE} than under Gavel (and \textit{Hadar} as well) for both clusters, due to the highest node resource utilization under \textit{HadarE} to let any training job be executed on multiple nodes concurrently and to leave no node idle, as long as job copies still exist for training.
\renewcommand{\arraystretch}{1.25}
\begin{table}[t]
\hfill
\fontsize{10pt}{12pt}\selectfont 
\caption{Comparative inference quality values for models trained under \textit{HadarE} versus under \textit{Hadar}}
\label{tab:inference comparison}
\vspace{20pt}
\resizebox{\columnwidth}{!}{
\begin{tabular}{|l|c|c|c|}
    \hline
    \textbf{Training Job}& \begin{tabular}[c]{@{}l@{}}\textbf{Forking} \\[-4pt] \textbf{(under \textit{HadarE})}\end{tabular} & \begin{tabular}[c]{@{}l@{}}\textbf{No Forking} \\[-4pt] \textbf{(under \textit{Hadar})}\end{tabular}  & \textbf{Metric} \\ \hline
    \begin{tabular}[c]{@{}l@{}}Language \\[-4pt] Translation \cite{transformer}\end{tabular} & 54.690 & 52.410  & ACC \\ \hline
    \begin{tabular}[c]{@{}l@{}}Image \\[-4pt] Classification \cite{resnet50}\end{tabular}   & 91.620 & 87.340 & ACC \\ \hline
    \begin{tabular}[c]{@{}l@{}}Recommendation \\[-4pt] System \cite{recoder}\end{tabular}   & 38.700  & 40.300  & MSE \\ \hline
    \begin{tabular}[c]{@{}l@{}}Language \\[-4pt] Modeling \cite{lstm}\end{tabular} & 4.310  & 4.460  & MSE \\ \hline
    \begin{tabular}[c]{@{}l@{}}MiMa Weather \\[-4pt] Predictions \cite{Mima} \cite{mima-codes} \end{tabular}   & 0.025 & 0.028  & MSE \\ \hline
\end{tabular}
}
\end{table}
\renewcommand{\arraystretch}{1}

\vspace{-10pt}
\textbf{Model Quality.} \textit{HadarE} accelerates training workload mixes and thus shortens average JCT substantially, by forking every DL training job (model) so that mutliple cluster nodes can train a given model concurrently for the highest CRU values. Amid training speedups, \textit{HadarE} is found to obtain trained models able to exhibit higher inference quality levels than those trained by \textit{Hadar}, as illustrated in Table~\ref{tab:inference comparison}. All the five DL models trained for workload mix M-5 under the two schedulers on our testbed cluster are compared for their inference quality results, in terms of accuracy (ACC) and mean squared error (MSE). The quality metric of ACC (or MSE) is for the language translation model \cite{transformer} and the image classification model \cite{resnet50} (or the remaining three models); see Table~\ref{tab:inference comparison}. It is found that the language translation model trained by \textit{HadarE} achieves higher accuracy at 54.69\%, in contrast to 52.41\%. Likewise, \textit{HadarE} trains the encoder-decoder transformer model to yield the smaller MSE of 0.025 versus 0.028 for the model trained by \textit{Hadar}. Interestingly, \textit{HadarE} trains all the five DL models to deliver higher quality than \textit{Hadar} consistently, despite its training speedups, possibly resulting from its use of heterogeneous cluster nodes to train models with more powerful ones undertaking larger number of steps before model parameter aggregation and consolidation (described in Section~\ref{subsec-aggregation}) for better model generality. Note that the DL models trained under other workload mixes (e.g., M-8, M-10, etc.) enjoy similar inference quality advantages by \textit{HadarE} than by \textit{Hadar}, so do the models trained on an AWS cluster.
\begin{figure*}[h]
    \centering
    \begin{minipage}[b]{0.49\textwidth}
        \centering
        \includegraphics[width=0.90\textwidth, height=0.42\textwidth]{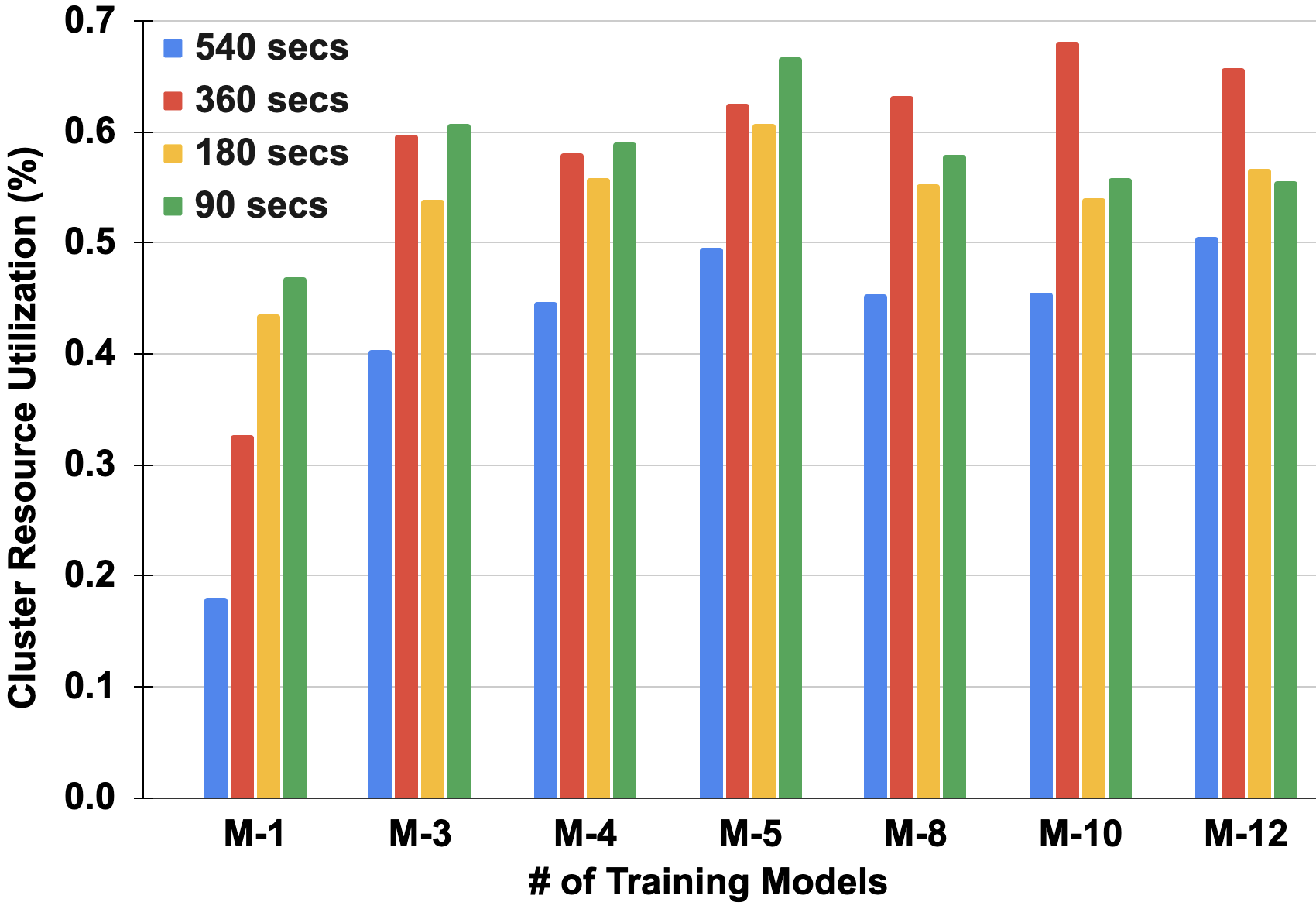}
        \caption*{(a) Results on AWS cluster}
        \vspace{15pt}
        \label{fig:irl-aws}
    \end{minipage}
    \hfill
    \begin{minipage}[b]{0.49\textwidth}
        \centering
        \includegraphics[width=0.90\textwidth, height=0.42\textwidth]{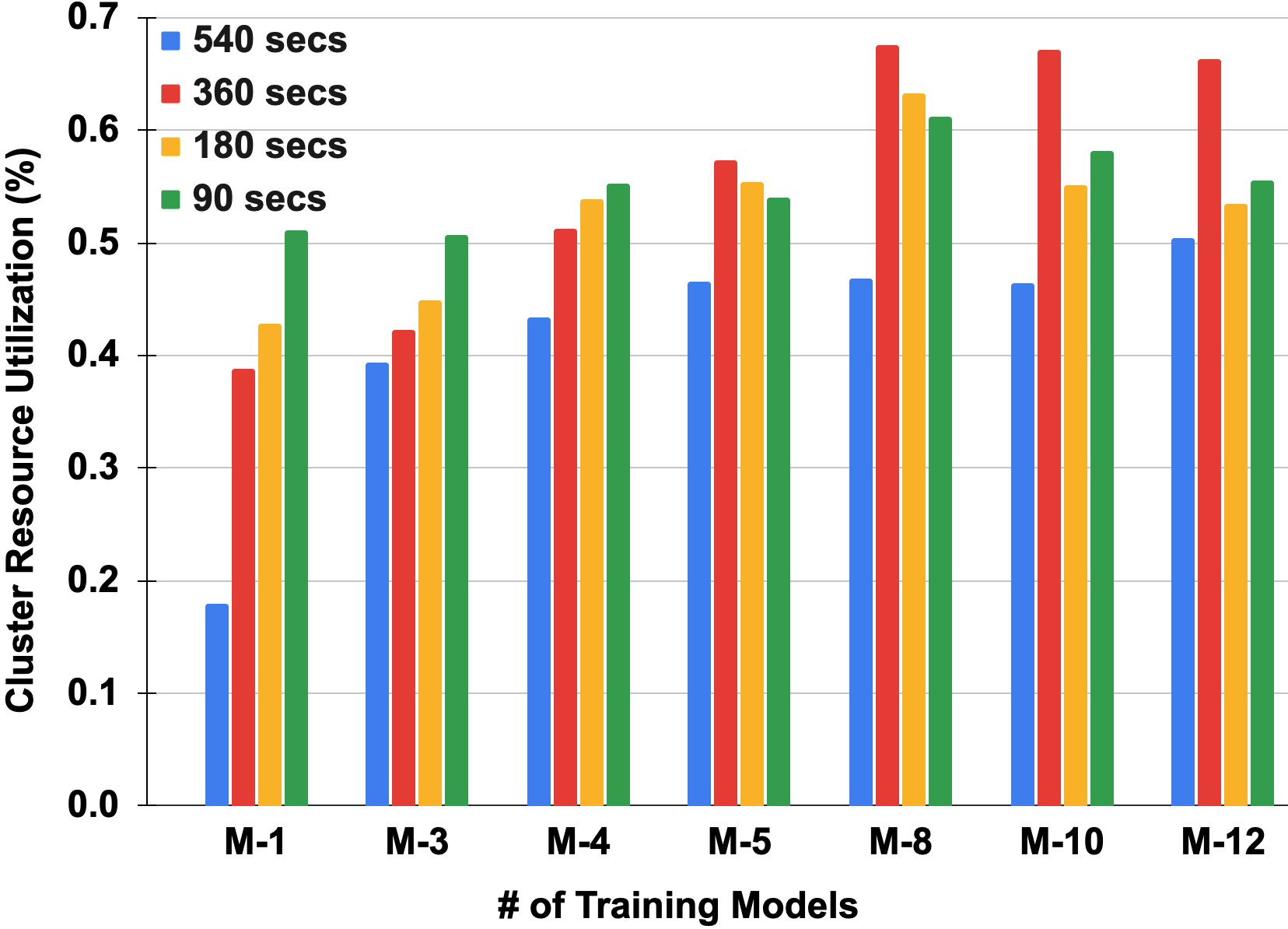}
        \caption*{(b) Results on testbed cluster}
        \vspace{15pt}
        \label{fig:irl-physical}
    \end{minipage}
\caption{CRU results of various slot time spans on AWS and testbed clusters under \textit{HadarE}.}
\vspace{0.25cm}
\label{fig:roundlength}
\end{figure*}
\vspace{10pt}
\begin{figure*}[h]
    \centering
    \begin{minipage}[b]{0.49\textwidth}
        \centering
        \includegraphics[width=0.90\textwidth, height=0.42\textwidth]{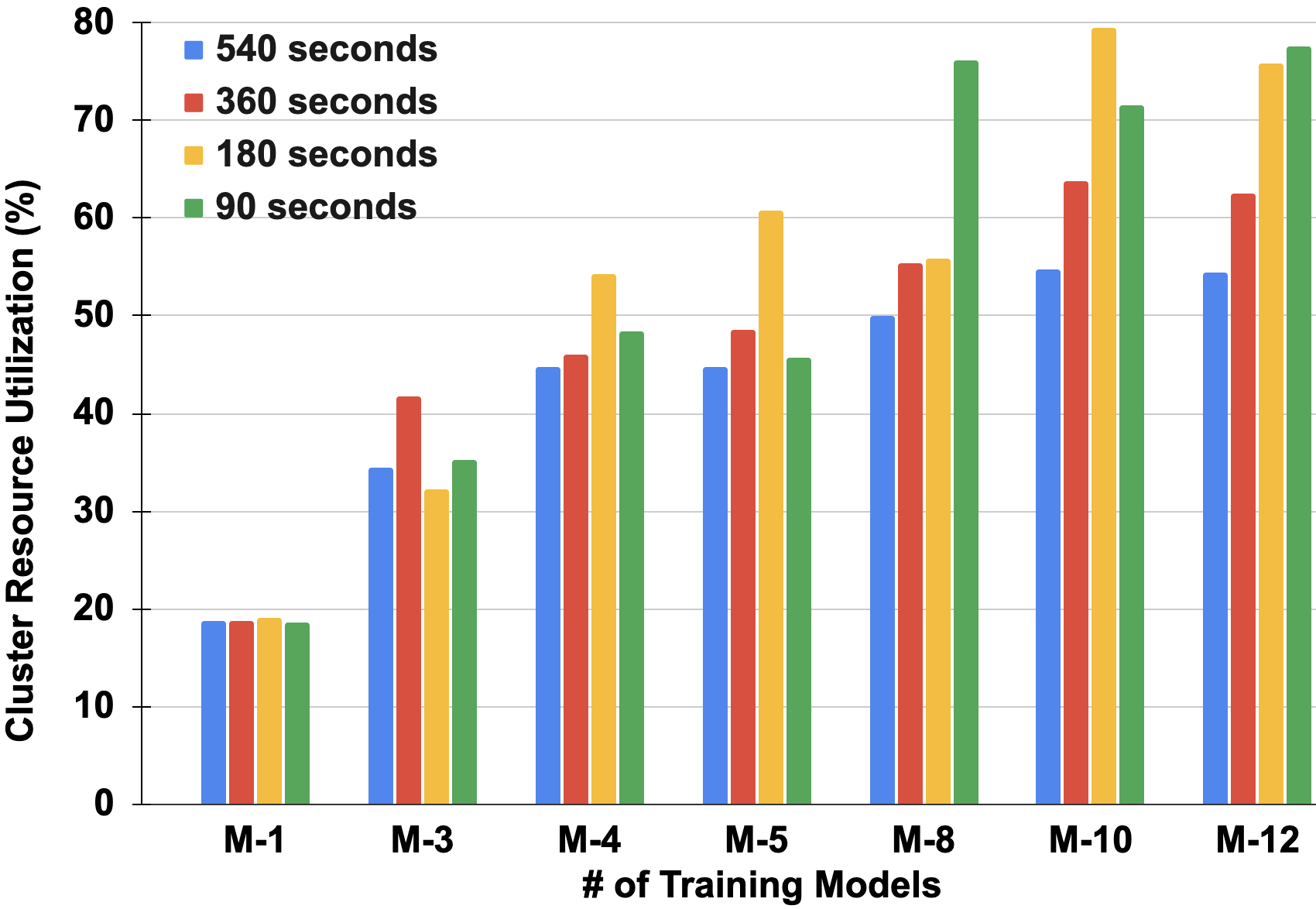}
        \caption*{(a) Results on AWS cluster}
        \vspace{15pt}
        \label{fig:irl-aws}
    \end{minipage}
    \hfill
    \begin{minipage}[b]{0.49\textwidth}
        \centering
        \includegraphics[width=0.90\textwidth, height=0.42\textwidth]{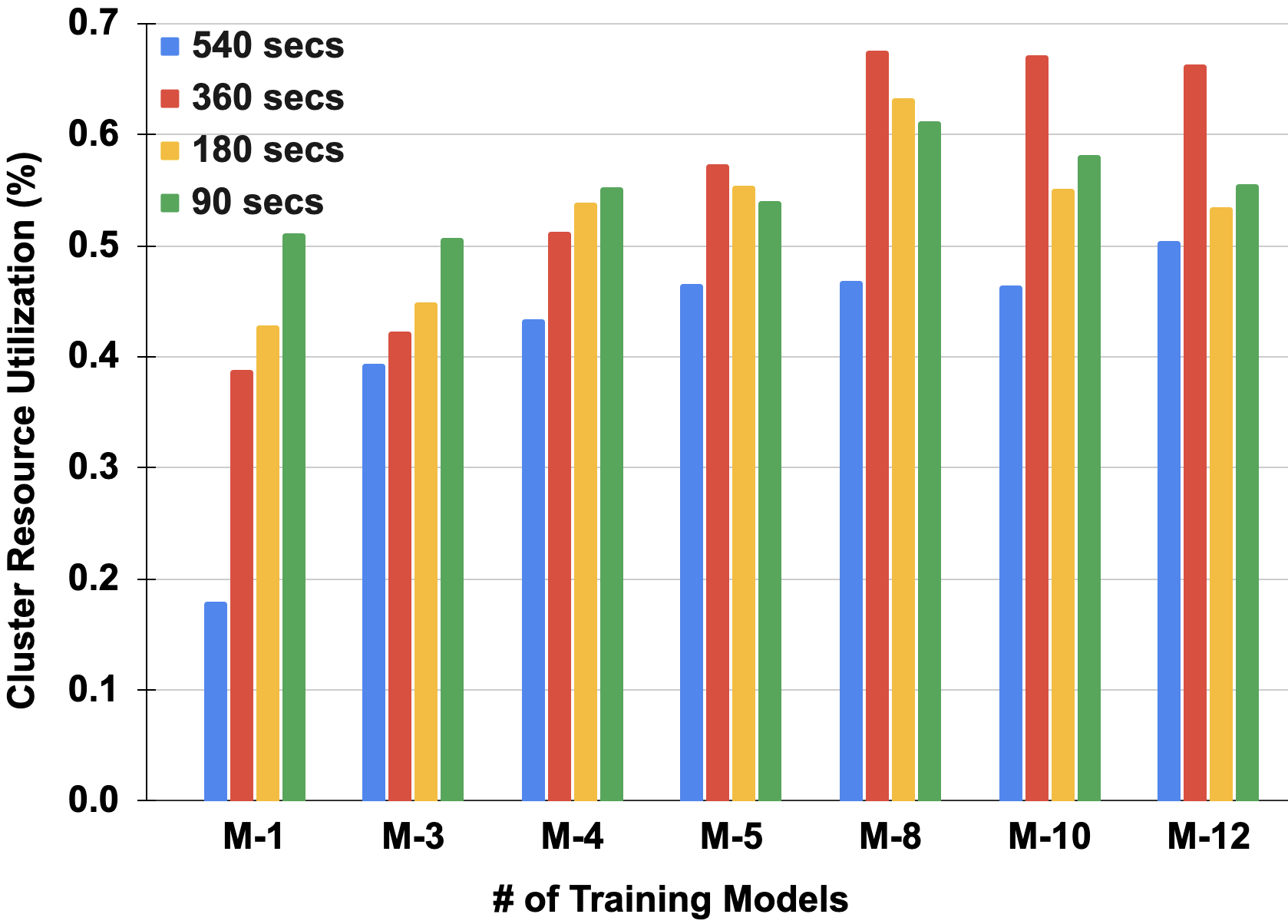}
        \caption*{(b) Results on testbed cluster}
        \vspace{15pt}
        \label{fig:irl-physical}
    \end{minipage}
\caption{CRU results of various slot time spans on AWS and testbed clusters under \textit{Hadar}.}
\label{fig:roundlength-hadar}
\end{figure*}
\vspace{-10pt}
\subsection{\textbf{Impact of Time Slots}}
Models are trained under \textit{HadarE} on rounds of a fixed slot time span, as under \textit{Hadar}. Upon completing the assigned number of training steps for a model or when the slot time expires, a cluster node notifies the Job Tracker (depicted in Fig.~\ref{fig:block-diagram}) of its obtained model parameter values and training progress in terms of the finished training step count, for model aggregation and consolidation therein before scheduling the next round of training job assignments, as detailed in Section \ref{subsec-aggregation}.
Intuitively, the training performance of a given workload mix is to be impacted by slot time length, with a smaller length to have better performance because the workload is then better distributed across all cluster nodes. However, \textit{HadarE} involves overhead due (1) mainly to communications between the Job Tracker and every assigned cluster node and (2) slightly to model aggregation and consolidation, making an excessively short slot time unfavorable. As a result, its best training performance (with the highest CRU) is expected to vary for different workload mixes, as demonstrated in Fig.~\ref{fig:roundlength}. From Fig.~\ref{fig:roundlength}(a) (or Fig.~\ref{fig:roundlength}(b)), training performance peaks at the slot time of 360 seconds for large workload mixes, namely, M-8 to M-12 on the AWS cluster (or M-5 to M-12 on our testbed cluster), as somewhat expected. When the slot time shrinks below 360 seconds, the overhead amounts then dwarf the benefits due to better workload distribution among cluster nodes. For small workload mixes, on the other hand, a short slot time yields the highest CRU, as can be observed for M-1 to M-5 (or M-1 to M-4), to enjoy the best training performance under the slot time of 90 seconds on the AWS cluster (or our testbed cluster).

Like \textit{HadarE}, \textit{Hadar} also incurs communication overhead between its Scheduler and assigned cluster nodes (see Fig.~\ref{fig-overview}), albeit at a lighter degree due to fewer jobs involved (without forking). Unlike \textit{HadarE}, it is free from model aggregation and consolidation. The CRU results of various slot time spans for workload mixes trained on the AWS cluster (or our testbed cluster) under \textit{Hadar} are depicted in Fig.~\ref{fig:roundlength-hadar}(a) (or Fig.~\ref{fig:roundlength-hadar}(b)). It is observed from Fig.~\ref{fig:roundlength-hadar}(a) that CRU results on an AWS cluster are largest under a short time slot of 90 or 180 seconds for all workload mixes, except M-3 (which peaks under the time slot of 360 seconds). This is expected because \textit{Hadar} incurs a lighter overhead than \textit{HadarE} to favor a shorter time slot. On our testbed cluster, whose three nodes (out of five) have relatively old motherboards with slow PCIe 3.0 to suffer from larger communication overhead, however, CRU results peak at a long slot time of 360 seconds for M-5 to M-12, as illustrated in Fig.~\ref{fig:roundlength-hadar}(b). For small workload mixes (of M-1 to M-4), however, the best training performance is achieved under the shortest slot time of 90 seconds partially because DL models can be trained mostly on cluster nodes with new motherboards to have PCIe 4.0 that curbs communication overhead, thus making better job workload distribution which outweigh increased communication overhead.

\section{\textbf{Conclusion}}
This paper has treated a novel task-level heterogeneity-aware cluster scheduler, \textit{Hadar}, which aims to optimize such performance metrics as cluster resource utilization, total time duration, and average job completion time. The novel scheduler is formulated into an optimization problem for its solution, utilizing the primal-dual framework for task-level resource allocation across both temporal and spatial dimensions. A theoretical analysis on the proposed scheduler has been undertaken to show its polynomial runtime and a long-term performance guarantee with a bounded competitive ratio for job utility, implying approximate optimal solutions within proven constant bounds. Leveraging the dynamic programming structure, \textit{Hadar} generates optimal scheduling decisions effectively. In addition, \textit{Hadar} is enhanced by forking each job into multiple copies to let jobs trained concurrently on heterogeneous GPUs resided on separate cluster nodes to further boost CRU levels and thus shorten the total time duration, arriving at \textit{HadarE}. Besides considerable training acceleration, \textit{HadarE} is also shown to train DL models with high inference quality than \textit{Hadar}.

\bibliographystyle{IEEEtran}
\renewcommand{\refname}{\textbf{References}}
\bibliography{11_references.bib}

\end{document}